\documentclass[12pt]{article}

\usepackage[table]{xcolor}
\usepackage{scicite}
\usepackage{lipsum}                     
\usepackage{xargs}    
\usepackage{booktabs}
\usepackage{graphicx}
\usepackage{epstopdf} 

\usepackage{amsmath}
\usepackage{multirow}

 \usepackage{xr}             

\makeatletter
\newcommand*{\addFileDependency}[1]{
\typeout{(#1)}
\@addtofilelist{#1}
\IfFileExists{#1}{}{\typeout{No file #1.}}
}
\makeatother

\usepackage{hyperref}       
\hypersetup{hidelinks}
\usepackage{adjustbox}
\usepackage{tcolorbox}
\usepackage{float}
\usepackage{url}
\usepackage{xcolor}  

\newcommand{\TabPopulationData}{Table~S1}

\newcommand{\TabTopicsByGroup}{Table~S3}
\newcommand{\TabTopicsUnified}{Table~S4}

\newcommand{\TabCEMCountries}{Table~S6}
\newcommand{\TabPTAttention}{Table~S7}
\newcommand{\TabPTFunding}{Table~S8}
\newcommand{\TabDiDFunding}{Table~S9}

\newcommand{\FigShareAttention}{Fig.~S1}
\newcommand{\FigDomForCount}{Fig.~S2}
\newcommand{\FigDisciplineRatio}{Fig.~S3}
\newcommand{\FigAttnVsOutput}{Fig.~S4}
\newcommand{\FigAttnVsGDP}{Fig.~S5}
\newcommand{\FigRankChanges}{Fig.~S6}
\newcommand{\FigASAffCountries}{Fig.~S7}
\newcommand{\FigASSubjectAreas}{Fig.~S8}
\newcommand{\FigTopicDynamics}{Fig.~S9}
\newcommand{\FigFundingVsGDP}{Fig.~S10}
\newcommand{\FigReligionMigration}{Fig.~S11}
\newcommand{\FigMigrationVsGDP}{Fig.~S12}
\newcommand{\FigPreMigVsAttn}{Fig.~S13}
\newcommand{\FigCEMBalance}{Fig.~S14}
\newcommand{\FigScopusQuery}{Fig.~S15}

\usepackage{times}

\newenvironment{sciabstract}{%
\begin{quote} \bf}
{\end{quote}}

\newcounter{lastnote}

\title{Arab Spring's Impact on Science through the Lens of Scholarly Attention, Funding, and Migration}

\author
{Yasaman Asgari$^{1,2\ast}$, Hongyu Zhou,$^{3,4}$ Özgür Kadir Özer$^{5}$,\\ Rezvaneh Rezapour$^{6}$, Mary Ellen Sloane$^{7}$, and Alexandre Bovet $^{1,2\ast}$ \\
\\
\normalsize{$^{1}$Department of Mathematical Modeling and Machine Learning (DM3L)}\\ \normalsize{University of Zurich, 8057 Zurich, Switzerland}\\
\normalsize{$^{2}$Digital Society Initiatives, University of Zurich, 8001 Zurich, Switzerland}\\
\normalsize{$^{3}$Department of Computer Science and Technology,}\\ \normalsize{ University of Cambridge, Cambridge, United Kingdom}\\
\normalsize{$^{4}$Centre for R\&D Monitoring (ECOOM), University of Antwerp, Antwerp, Belgium}\\
\normalsize{$^{5}$Department of Science and Technology Policy Studies (TEKPOL)}\\\normalsize{Middle East Technical University, Ankara, Türkiye}\\
\normalsize{$^{6}$Department of Information Science, Drexel University, Philadelphia, United States}\\
\normalsize{$^{7}$Tennessee 
STEM Education Center, Middle Tennessee State University,}\\\normalsize{ Murfreesboro, United States}\\
\normalsize{$^\ast$ \{Yasaman.asgari, Alexandre.bovet\}@uzh.ch}
}

\date{}

\begin{document} 


\baselineskip24pt


\maketitle



\begin{sciabstract}
The 2010–2011 Arab Spring reverberated far beyond politics, reshaping how the Middle East and North Africa region (MENA) is studied. Analyzing 3.7 million Scopus-indexed articles published between 2002 and 2019, we find that mentions of ten of these countries in titles or abstracts rose significantly after 2011 relative to the global baseline, with Egypt receiving the greatest attention in the region. We link this surge to two intertwined mechanisms: an increase in research funding directed at the MENA region and the emigration of researchers who continued publishing on their countries of origin. Our analysis reveals that Saudi Arabia has emerged as a regional hub for studying the affected countries, attracting funding and scholars, and thereby playing a significant role in shaping the scientific narrative on the region. These findings demonstrate how political upheaval can reshape global knowledge flows by altering who studies whom, with what resources, and in which disciplines.
\end{sciabstract}

Science and research do not happen on their own; they are shaped by and help shape the economic and sociopolitical environment of their societies \cite{seth2009putting,lee2011research,miao2022latent}. Major events, such as revolutions, armed conflicts, and natural disasters, profoundly disrupt scientific ecosystems, resulting in reduced research productivity, weakened international collaboration, and increased brain drain \cite{zhang2024influence, ganguli2014scientific, derassenfosse2023effects, jin2004long, okolo2023economics, waldinger2016bombs}. The Arab Spring, sparked by Mohamed Bouazizi’s self-immolation in Tunisia in 2010, exemplifies this phenomenon across the Middle East and North Africa (MENA), affecting nearly every country in the region \cite{tufekci2012social}.

To date, investigations into the Arab Spring’s effects on scholarly ecosystems have relied mainly on bibliometric analysis with a focus on collaboration patterns \cite{zewail_dire_2014, ahmad_bibliometric_2021, vera_axyonova_academics_2022}, academic freedom \cite{saliba_academic_2018, vera_axyonova_academics_2022}, and research productivity \cite{ibrahim_arab_2018, cavacini_recent_2016, al-kindi_cardiovascular_2015}. More recent works in political science, international relations, and sociology have shown a marked surge in academic discourse on the transformations of the Middle East and North Africa (MENA) following the uprisings \cite{almaghlouth2015frames, cammett2021political}. 

However, there remains a notable absence of large-scale, systematic analyses that (1) identify who is studying which MENA countries, (2) trace how this attention varies across disciplines, and (3) reveal the evolution of this attention over time. Here, we address this gap by defining \emph{scholarly attention} as the frequency with which countries are mentioned in article titles and abstracts, and take this as a first lens to understand the effect of the Arab Spring on the geographic distribution of research focus. Country naming practices have already proven effective in capturing global research patterns in domains ranging from public health \cite{balsamo2025public} and the social sciences \cite{castro2022north, castro2025revisiting} to research on the United Nations' Sustainable Development Goals (SDGs) \cite{henkel2023studies}. 

A second lens is research funding: naturally, research output is primarily a function of the funding available; thus, it is essential to track the origins of funding. Indeed, funding plays a decisive role in determining the scale, thematic focus, and sustainability of research activities \cite{glaser2018changing, lepori2007comparing, lepori2019scientific, heyard2021value, hussinger2022long, zhang2018joint, mosleh2022scientific, sattari2022ripple}. In developing countries, which often have limited domestic financial resources, foreign funding plays a more pronounced role in research activities compared to those in the developed economies \cite{alvarez2018studying,chankseliani2023funds,el-ouahi2024research}. While foreign funding enhances collaboration and citation impact, it can also create dependency and influence on research agendas \cite{chinchilla2019follow, glaser2018changing, nkansah2024dependency, sabzalieva2020negotiating}.

A third lens is scholarly migration: when researchers relocate, they often continue research on their home country, sustaining or redirecting attention from abroad.  Bibliometric data have been recently used to track the movement of researchers across borders, revealing important migration patterns \cite{machavcek2022researchers,akbaritabar2024bilateral,Zhao2022, Zhao2021, Miranda-Gonzalez2020,momeni_many_2022} and the key factors driving them \cite{Zhao2021, Sanliturk2023, Zhao2023, Akbaritabar2023, momeni_many_2022}. Using long-term trends helps scholars to study how major socio-political events shape brain drain and brain gain dynamics, for instance during World War II \cite{waldinger2016bombs}, in China’s Cultural Revolution \cite{jin2004long}, the collapse of the Soviet Union \cite{ganguli2014scientific, Subbotin2021}, Brexit \cite{Sanliturk2024}, and the Ukraine war \cite{derassenfosse2023effects}.

Together, these three lenses---attention, funding, and migration---offer a holistic view of how the Arab Spring reshaped global research dynamics. Moreover, examining the link between attention and both funding flows and scholarly migration gives a unique perspective on the forces shaping these shifts. Here, we present a longitudinal analysis of scholarly attention towards ten countries affected by the Arab Spring, drawing on Ibrahim’s 2018 classification of sociopolitical trajectories into government overthrows (Egypt, Tunisia), civil wars (Libya, Syria, Yemen), and governmental changes (Bahrain, Jordan, Kuwait, Morocco, Oman), as well as major and minor protests\cite{ibrahim_arab_2018}. 

We extracted mentions of countries from the titles and abstracts of over 25 million articles published between 2002 and 2019 and identified 73,000 papers that mention at least one of the target countries. As a comparison baseline, we also analyzed 3.7 million articles that mention any country. Moreover, we obtained funding information embedded in acknowledgments and used a scholarly migration database \cite{ScholarlyMigrationDatabase2022} to examine the relationship between changes in scholarly attention, funding, and migration. We employed a difference-in-differences approach to estimate the causal impact of the Arab Spring on scholarly attention.

Our findings indicate that the target countries experienced a statistically significant increase in scholarly attention compared to the rest of the world. Within this group, Egypt emerged as the primary focus of research, with its post-uprising attention growth exceeding 80\% of all the countries globally. Attention patterns differed by researcher affiliation: foreign scholars disproportionately contributed to social science publications, whereas domestic scholars focused more on the health and life sciences. Foreign-funded research increased significantly, primarily from Saudi Arabia, the United States, and the United Kingdom. The social sciences, although historically underfunded, have seen increased funding, which has fueled research growth in this field. 

Regarding scholarly migration, Egypt remained a research hub, attracting scholars, while Tunisia and Syria experienced severe brain drain as researchers relocated to wealthier nations. Western and Islamic countries, especially Saudi Arabia, Qatar, the UAE, and Malaysia, emerged as top destinations, presumably due to economic opportunities and cultural affinities. The strong correlation between scholarly migration and scholarly attention suggests that relocated researchers continued studying their home countries, contributing to increased foreign attention.

These findings underscore the Arab Spring’s lasting impact on academia and reveal the power dynamics between sources and targets of scholarly attention. The relationship between scholarly attention, funding, and scholarly migration demonstrates how major socio-political events reshape scientific inquiry. Our study provides a foundational framework for understanding these dynamics, offering insights applicable to other regions undergoing major transformations.

\section*{Results}
\subsection*{Shifts in scholarly attention before and after Arab Spring}
\begin{figure}[!ht]
    \centering
    \includegraphics[width=\textwidth]{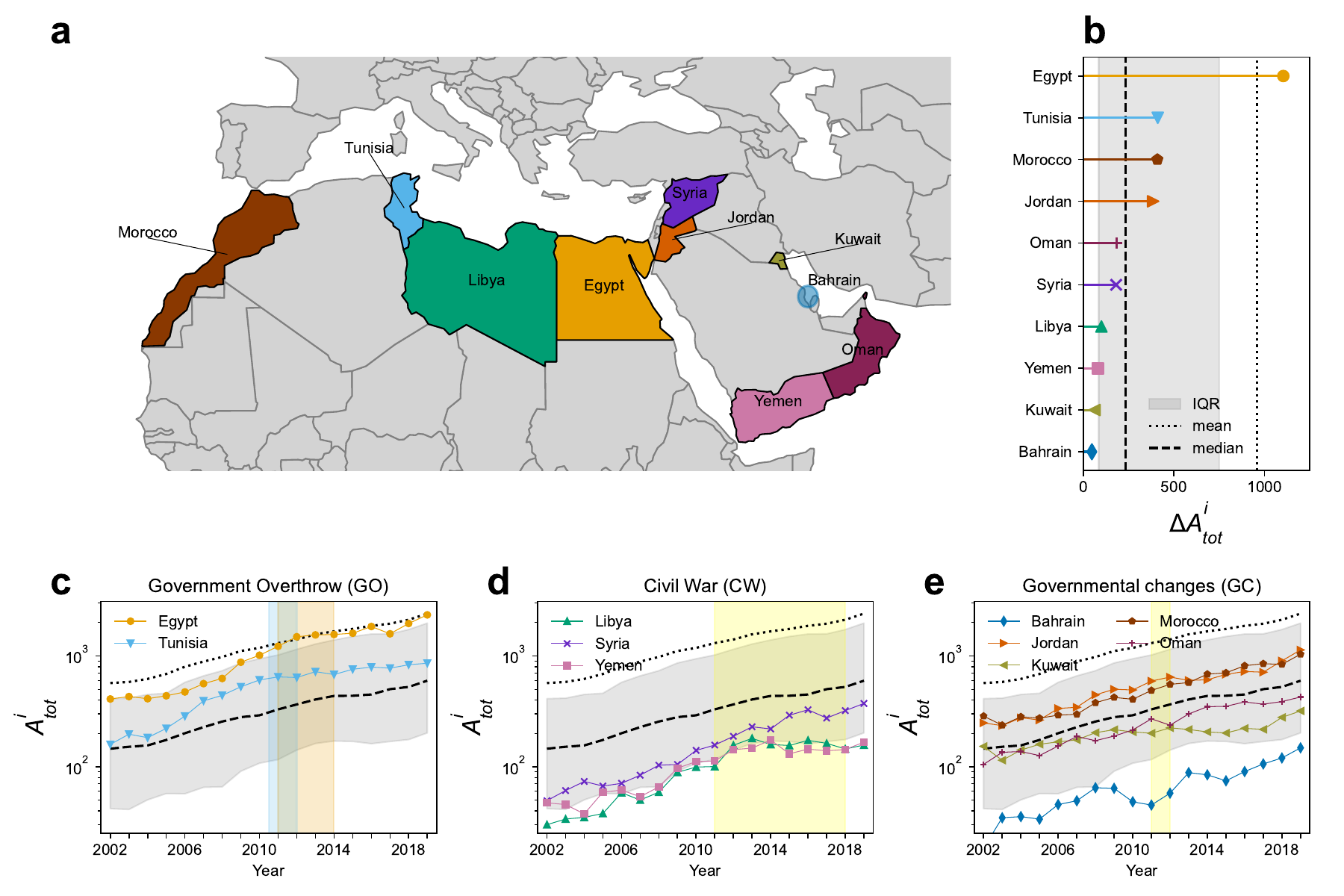}
  \caption{\textbf{Scholarly Attention in Target Countries Pre- and Post-Arab Spring.} (\textbf{a}), Geographical locations of the target countries. (\textbf{b}), The difference in total scholarly attention before and after the Arab Spring across target countries highlights non-uniform changes. In all panels, the gray area represents the interquartile range for global trends, with the dashed line indicating the median and the dotted line showing the mean. Country-specific trends reveal significant increases in the volume and distribution percentiles of scientific publications for countries experiencing government overthrow (\textbf{c}), civil war (\textbf{d}), and governmental changes (\textbf{e}). The vertical spans indicate periods of unrest in these countries associated with the Arab Spring and its aftermath.}
    \label{fig_attention_temporal_trend} 
\end{figure}

The target countries in this study are located in North Africa or the Middle East (Figure \ref{fig_attention_temporal_trend}a). These countries range from highly populous nations, such as Egypt, to less populous ones, including Bahrain and Kuwait (see \TabPopulationData{} in the Supplementary Materials). 

We quantify the scholarly attention these countries received via their mentions in Scopus publication titles and abstracts, which we represent as a directed, weighted network: each node corresponds to a country, and a directed edge from country $j$ to country $i$  fractionally counted number of papers authored by researchers at institutions in $j$ that mention $i$ (see the Materials and Methods section for a detailed explanation). A country’s total received attention is given by $A^i_{\mathrm{tot}} = \sum_{j} A^i_j$, and temporal change in attention is computed as the difference in its average annual attention between the pre-Arab Spring period (2002–2010) and the post-Arab Spring period (2011–2019).

Figure \ref{fig_attention_temporal_trend}b highlights that shifts in scholarly attention towards each country \(i\), namely \(\Delta A^i_{tot}\), are not evenly distributed across the countries. Egypt surpasses the third quartile and the mean, whereas Tunisia, Morocco, and Jordan fall between the median and the third quartile. Meanwhile, the remaining countries are positioned on and below the median. 

While the attention temporal trajectories differ across these countries, all of them received more annual scholarly attention in 2019 than in 2002 (Fig. \ref {fig_attention_temporal_trend}c–e). Research on Egypt began accelerating after 2009, then rose above the global average between 2011 and 2015, dipped slightly in 2016, and then stabilized around the world's mean (Fig. \ref{fig_attention_temporal_trend}c). Attention towards Tunisia climbed steadily from 2005 until about 2010, after which it hovered at a stable plateau. In the civil war group (Fig. \ref{fig_attention_temporal_trend}d), Yemen and Libya received minimal attention until 2009, followed by a surge that plateaued until 2012; Libya reached its peak in attention in 2013, while attention towards Yemen peaked in 2014, before both declined. Syria experienced a steady rise in attention, with a peak in 2016 probably corresponding to the rise of ISIS and associated humanitarian crises. Among the governmental changes group, attention towards Jordan and Morocco rose above the global median consistently; attention directed at Oman experienced a brief decline in 2012; attention towards Kuwait remained relatively flat; and Bahrain, despite a low overall volume, showed peaks in attention in 2008 and 2013, followed by growth after 2015. 

When normalizing each country’s attention relative to the total global attention, the fraction of attention targeted at most countries displays stable trajectories, except for Egypt, Tunisia, and Syria, which show notable increases (see \FigShareAttention{} in the Supplementary Materials).

\subsection*{Share of Arab Spring-related research}
One possible explanation for the rise in scholarly attention is the increased research related to the Arab Spring. To assess how much of this attention is driven explicitly by the Arab Spring's aftermath, we investigate articles containing mentions of keywords related to the ``Arab Spring'' (e.g., Arab Uprisings, Arab Revolution, etc., see the Materials and Methods section for a comprehensive list and explanation of the methodology) in their title or abstract. 

Arab Spring–related research is concentrated primarily within the Social Sciences category (refer to Supplementary Materials \FigASSubjectAreas), in which the Social Sciences discipline accounts for 57.1\% of the publications. The second-largest share belongs to Arts and Humanities (17.6\%), followed by disciplines such as Economics, Econometrics and Finance (5.6\%), and Business, Management and Accounting (3.7\%), although their shares are considerably smaller.

\begin{figure}[!ht]
    \centering
    \includegraphics[width=\linewidth]{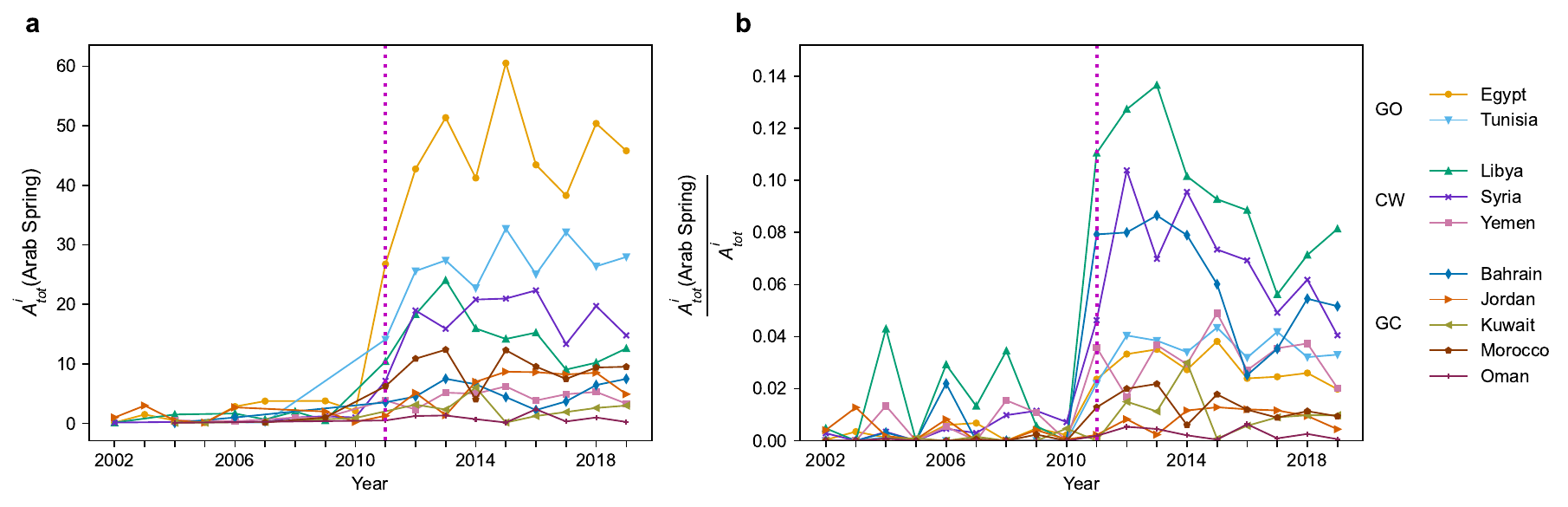}
\caption{\textbf{Scholarly attention related to research on the Arab Spring.} (\textbf{a}) Country-specific scholarly attention related to research on the Arab Spring.
Research focusing on the government overthrow (GO) group attracted the highest scholarly attention, followed by the civil war group (CW). (\textbf{b}), Proportion of Arab Spring-related research as part of total scholarly attention by target country. Syria and Libya, classified under the civil war group, exhibit the highest shares, whereas the governmental changes group shows minimal shares. The vertical dotted line represents the year 2011, marking the onset of the Arab Spring.}
    \label{arab_spring_related_research}
\end{figure}

In Figure \ref {arab_spring_related_research}, we show that only a small fraction of the scholarly attention to a target country following the Arab Spring can be credited to Arab Spring-related research. The government-overthrown group, specifically Egypt, attracted the most attention in Arab Spring-related research, followed by countries in the civil war group (Fig. \ref{arab_spring_related_research}a). The proportion of scholarly attention related to the Arab Spring, compared to the total attention received, reveals that the civil war group holds a significantly larger share than the others(Fig. \ref{arab_spring_related_research}b). The government overthrow group demonstrates relatively steady attention share around 3-4\%, and for the governmental changes group, except Bahrain, the share is generally below 2\%. Libya recorded the highest share in 2013, accounting for almost 14\%. Syria shows two notable peaks, one in 2012 and another in 2014. Bahrain's presence among the top three in terms of share is due to its low total attention.

However, the temporal dynamics of scholarly attention in Arab Spring–related research not only vary across different political trajectories but also reveal distinct thematic emphases. To capture these differences, we applied BERTopic for topic modeling on the retrieved documents (see Supplementary Materials: \FigTopicDynamics{}, supplementary text, \TabTopicsByGroup{}, and \TabTopicsUnified. 
Social media emerges as the most prominent theme across groups. In cases of government overthrow, the number of articles reached its peak in 2013. In the civil war, it reached its peak in 2015 and then decreased. For governmental changes, the rate has been growing since 2017. Social media played a dual role in the Arab Spring, facilitating collective action by enabling protest coordination while also shaping political debates and public discourse within a new media environment \cite{howard2011opening, soengas2013role, movements2020impact}. 

Topics such as Islamist parties, women’s rights, and financial and oil markets appeared predominantly in the government overthrow group, with only minimal presence in the other groups. By contrast, Turkey’s regional role and refugees were concentrated in the civil war group, reflecting how instability from civil conflicts reverberated into concerns over regional spillovers and broader geopolitical risks \cite{chatty2017syrian}. Meanwhile, the theme of Gulf monarchies was characteristic of the governmental changes group, where references to Saudi Arabia and the Gulf Cooperation Council underscored the region's geopolitical influence in shaping how the Arab Spring was studied within these countries.

\subsection*{Differences between domestic and foreign scholarly attention across countries and disciplines}

We now turn to identifying the sources of scholarly attention toward the target countries. Scholarly attention to a country can come from two sources: researchers affiliated with that country (\textit{domestic attention}), namely \(A_i^i\) (self-loops in the scholarly attention network), and researchers from other countries (\textit{foreign attention}), namely \(\sum_{j \neq i }A_j^i\). 

\begin{figure}[!ht]
    \centering
    \includegraphics[width=\textwidth]{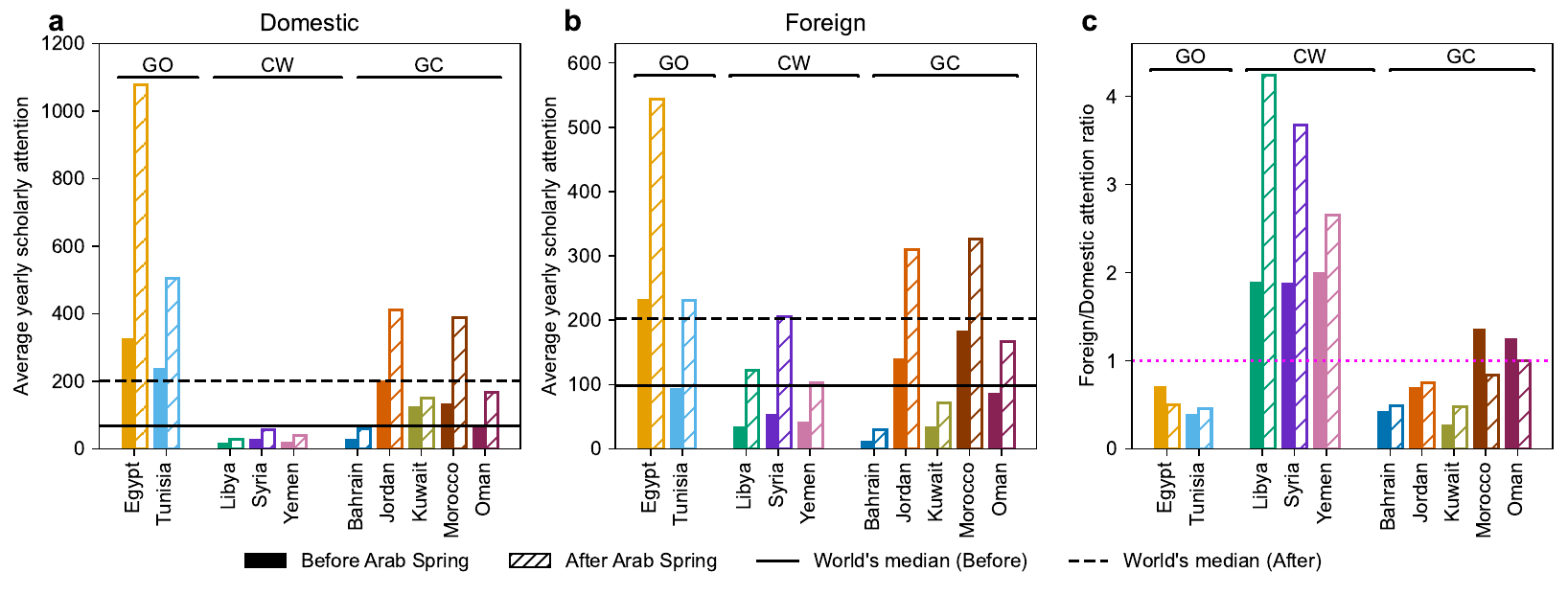}
     \caption{\textbf{Domestic and foreign scholarly attention towards target countries before (2002–2010) and after (2011–2019) the Arab Spring.} Attention from domestic (\textbf{a}) and foreign sources (\textbf{b}) and the foreign-to-domestic ratio (\textbf{c}) are shown for each country across three groups: government overthrow (GO), civil war (CW), and governmental changes (GC). Filled bars represent values before the Arab Spring, while hatched bars indicate values after the Arab Spring. The world's median attention before and after the Arab Spring is marked with solid and dashed black lines, respectively. }
    \label{domestic_vs_foreign_attention}
\end{figure}

Figure \ref{domestic_vs_foreign_attention}a shows that domestic scholarly attention increased across all target countries following the uprisings, though with substantial variation. The largest increase occurred in the government-overthrown group, which consistently received a higher level of domestic attention. This increase may reflect a revitalized local research environment supported by established universities and national rebuilding efforts. Meanwhile, countries in the civil war group showed only modest gains in domestic attention, a pattern likely influenced by prolonged conflict, institutional disruption, reduced research capacity, and large-scale scholarly migration. The governmental change group also experienced notable increases, particularly in Jordan and Morocco, where internal reforms may have contributed to a more stable environment for domestic scholarly activity.

Similarly, while foreign scholarly attention increased for all three groups (Fig. \ref{domestic_vs_foreign_attention}b), the increase in the government overthrow group is significant, surpassing the rise for the world's median. In the civil war group, Syria, notably, transitioned from being below the median before the Arab Spring to above it afterward. 

The foreign-to-domestic attention ratio (Fig. \ref{domestic_vs_foreign_attention}c) quantifies the relative contribution of foreign versus domestic scholarly attention for each country. In Egypt, the ratio declined to approximately 0.5 after the Arab Spring, indicating that domestic attention outpaced foreign attention. Oman and Morocco both transitioned from foreign-dominated attention (ratios above 1) to more balanced ratios, close to one. This shift may reflect an increase in domestic research capacity or institutional support. By contrast, the civil war group had ratios above one before the uprisings, which rose even further afterward, signaling a growing dominance of foreign over domestic attention.

These patterns show that domestic and foreign scholarly attention did not evolve uniformly across cases. Countries with more stable political transitions saw relatively greater domestic engagement, while those experiencing sustained conflict attracted increasing levels of international attention. Higher foreign-to-domestic attention ratios may be associated with international concern over conflict spillovers or humanitarian crises.

\begin{figure}
    \centering
\includegraphics[width=\linewidth]{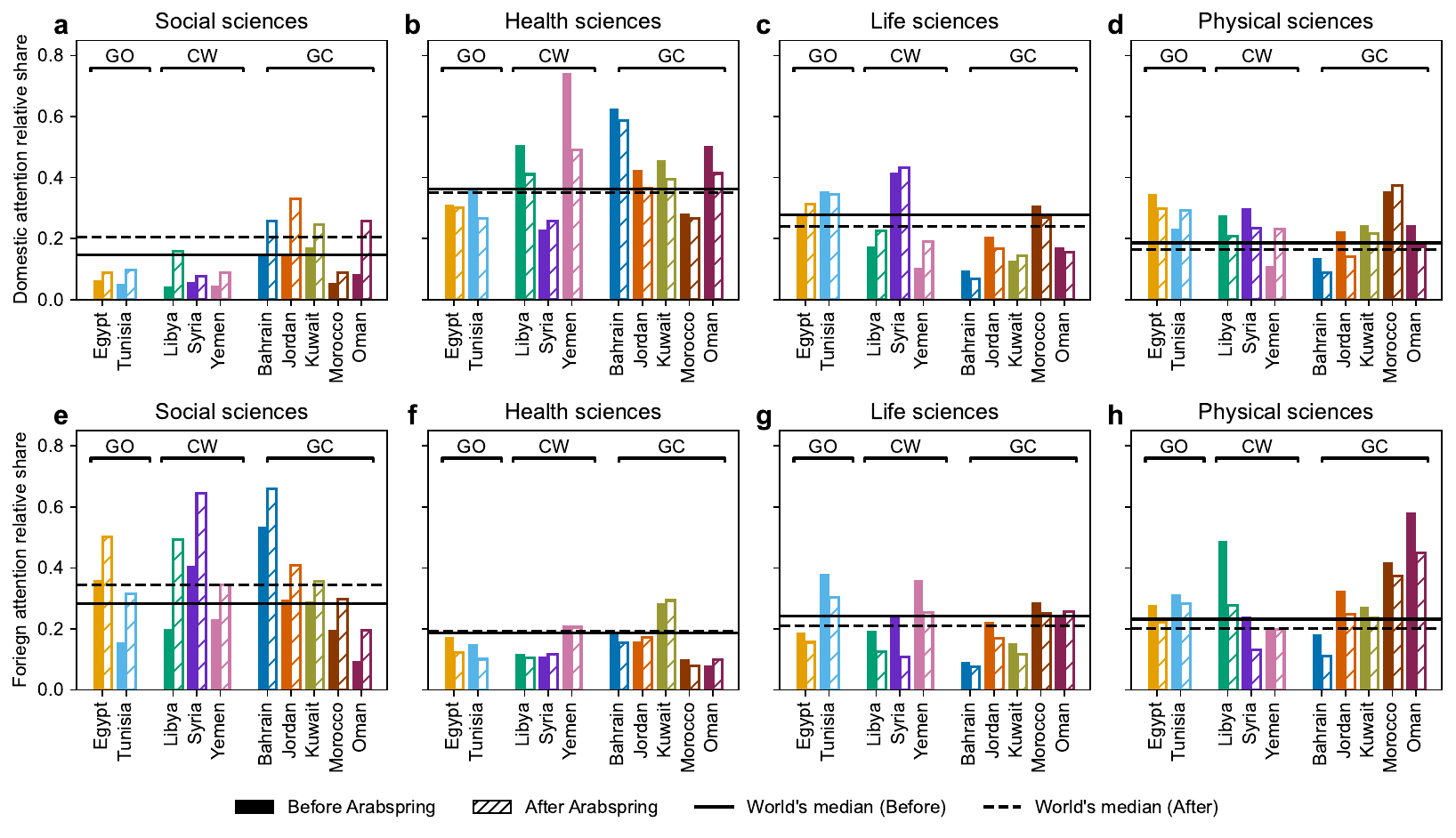}
    \caption{\textbf{Share of domestic and foreign scholarly attention before and after the Arab Spring by discipline.} The target countries are split into three groups: government overthrow (GO), civil war (CW), and governmental changes (GC). Filled bars represent values before the Arab Spring, while hatched bars indicate values after the Arab Spring. The world's median before and after the Arab Spring is marked with solid and dashed black lines, respectively. In domestic attention (\textbf{a–d}), social sciences consistently held the smallest share for most countries. After the Arab Spring, the proportion of social sciences in the governmental changes group (excluding Morocco) rose above the median. In foreign attention (\textbf{e–h}), social sciences showed a notably higher share, particularly in Yemen and Libya, where it became the dominant field post-Arab Spring.}
    \label{discipline_domestic_foriegn}
\end{figure}

Figure \ref{discipline_domestic_foriegn} shows how foreign and domestic attention differ across four major disciplines: social sciences, health sciences, life sciences, and physical sciences.
In domestic attention (Fig. \ref{discipline_domestic_foriegn}a-d), the relative shares of each discipline show that in the government overthrow and civil war groups, as well as Morocco, the social sciences consistently held the smallest share of attention before and after the Arab Spring. Although the other groups did not show notable shifts, after the Arab Spring, the proportion of social sciences in the government changes group (excluding Morocco) rose above the median. This shift could be attributed to government reforms, which may have reallocated attention toward social sciences. Meanwhile, the relative share of domestic scholarly attention to health sciences in Yemen has declined significantly following the Arab Spring, due to an increase in domestic attention to other disciplines (for analysis of the counts, refer to Supplementary Materials, \FigDomForCount). In foreign attention (Fig. \ref{discipline_domestic_foriegn}e-h), the social sciences consistently exhibit a much higher proportion than what is observed in domestic attention. In Yemen and Libya, the social sciences, which had previously been below the median before the Arab Spring, rose above the median afterward and became the dominant discipline. Moreover, in Morocco and Oman, the physical sciences receive the largest share of foreign attention, possibly driven by the regions’ historical and geographical research appeal.

We also computed the foreign-to-domestic attention ratio across individual disciplines (see \FigDisciplineRatio{} in the Supplementary Material). The decrease in this ratio for Egypt occurs across all disciplines, indicating a consistent pattern of increased domestic attention following the Arab Spring. In the civil war group, the social sciences have extremely high foreign-to-domestic ratios, highlighting a heavy reliance on foreign sources. Similarly, in the health sciences, foreign-to-domestic ratios rose, shifting from below one to above one. In the governmental changes group, except for Kuwait, the foreign-to-domestic attention ratio in social sciences decreased, showing a greater increase in domestic attention. These diverse patterns suggest a strong dependence of scholarly attention on political trajectories.

\subsection*{Main sources of the foreign increase in scholarly attention}

\begin{figure}[!ht]
    \centering
    \includegraphics[width=\textwidth]{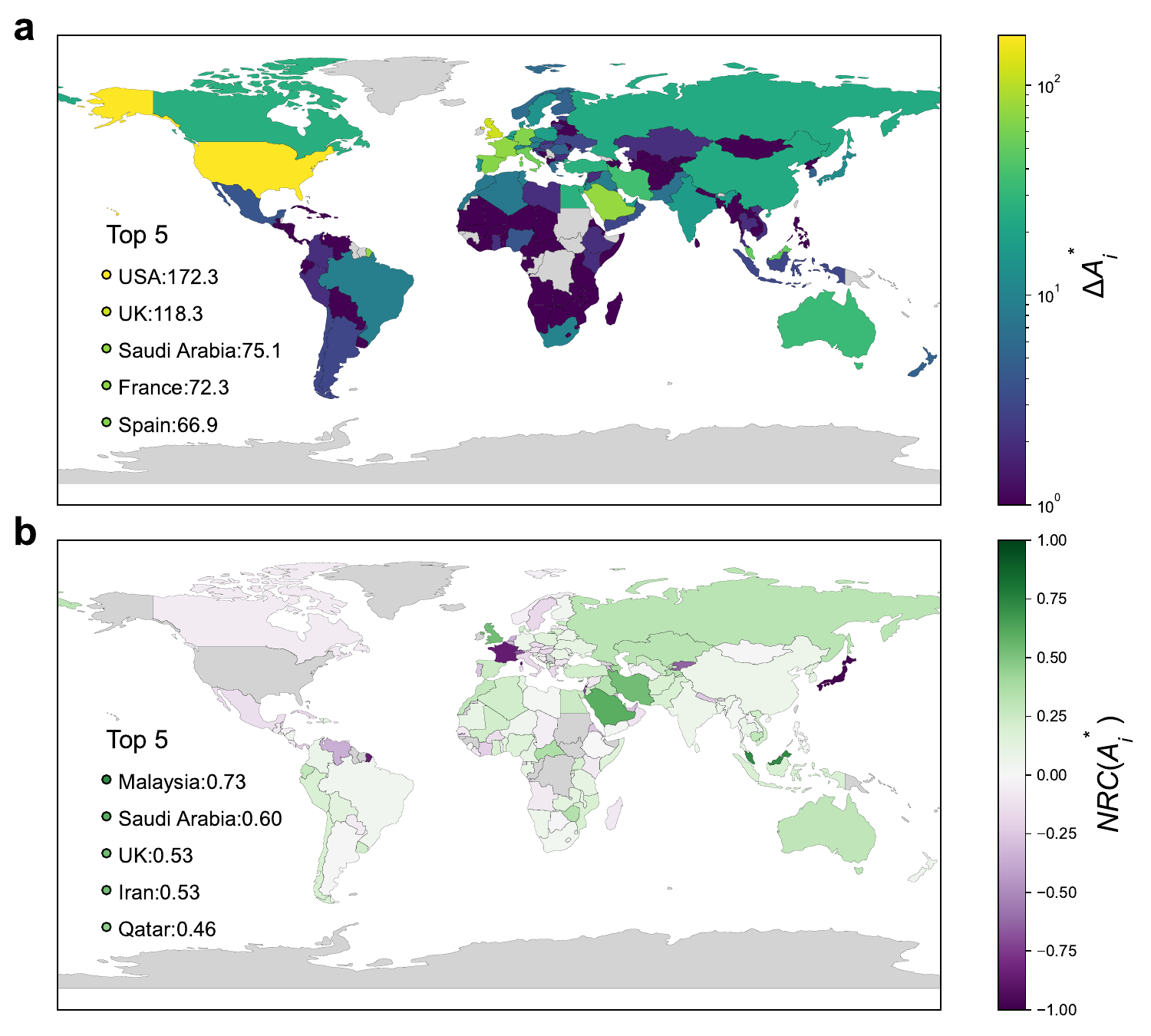}
    \caption{\textbf{Geographical distribution of foreign scholarly attention sources towards target countries}: (\textbf{a}) Differences in average yearly scholarly attention towards all target countries: Western countries alongside Saudi Arabia show the highest differences, while countries in the Global South show the least. (\textbf{b}) Normalized rank change capturing relative attention changes: The impact of the Arab Spring is more evident in the region and among Islamic countries. Saudi Arabia is among the top 5 countries in terms of the difference in average and rank inertia metrics, highlighting its significant influence in research on the target countries. Territories are visualized with Geopandas \cite{den2024geopandas}.} 
    \label{geo_attention}
\end{figure}

\begin{figure}[!ht]
    \centering
    \includegraphics[width=\textwidth]{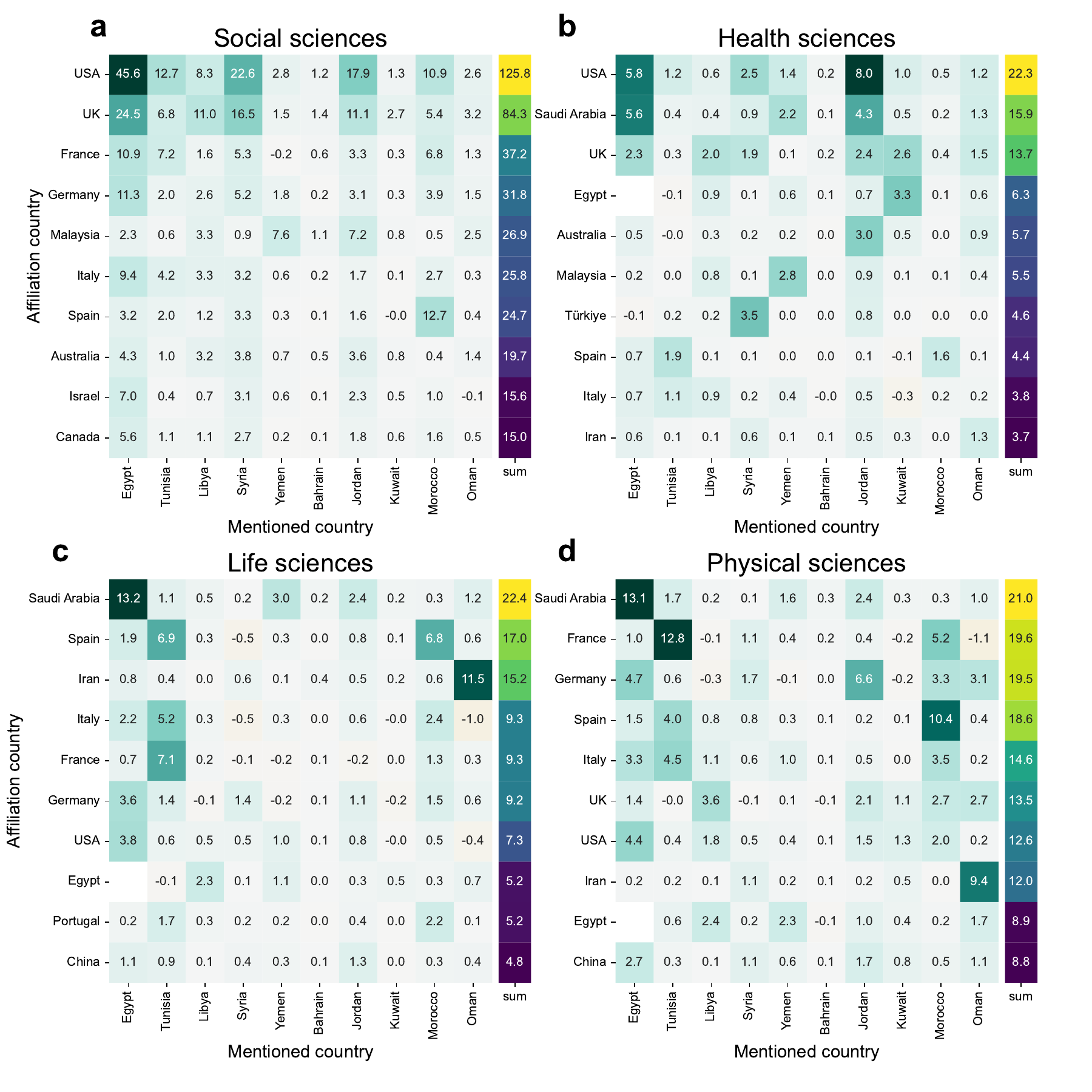}
    \caption{\textbf{Repartition of scholarly attention towards target countries by discipline and source country}: Differences in yearly scholarly attention by country, listing the top 10 countries with the greatest differences in the field of Social sciences (\textbf{a}), Health Sciences (\textbf{b}), Life Sciences (\textbf{c}), and Physical Sciences (\textbf{d}).}
    \label{top_ten_attention_discipline}
\end{figure}

We now examine the geographic sources of foreign scholarly attention (see Figs. \ref{geo_attention} and \ref{top_ten_attention_discipline}) to identify which regions and countries contributed to the observed increase in foreign attention toward Arab Spring-affected nations.

The contributions of different countries to the rise in foreign scholarly attention show significant geographical disparities (Fig. \ref{geo_attention}a). Western nations, known for their high scholarly output, account for the largest share of this growth, whereas contributions from the Global South remain relatively limited. The United States leads, with an average increase of (+172.3  papers/year), followed by the United Kingdom (+ 118.3 papers/year). Malaysia and Saudi Arabia also contribute significantly despite their lower scholarly output. 

The increase in scholarly attention of sources countries strongly correlates with their total scholarly output during the period (Pearson correlation: 0.684; Spearman correlation: 0.845, both with \(p < 0.0001\), see \FigAttnVsOutput{} in the Supplementary Materials) and with their average GDP during this period (Pearson correlation: 0.432; Spearman correlation 0.608, both with \(p < 0.0001\), see \FigAttnVsGDP{} in the Supplementary Materials) and with pre–Arab Spring yearly average out-migration from target countries (Pearson correlation: 0.7813; Spearman correlation: 0.7453, both with \(p < 0.0001\), see \FigPreMigVsAttn{} in the Supplementary Materials). These three factors--GDP, scholarly output, and pre–Arab Spring migration flows--possibly explain these disparities.

To understand the effect of the Arab Spring, we examine the changes in the positions of target countries within the rankings of countries that are sources of attention. We show the change in ranks of target countries for the 10 most important sources of attention in \FigRankChanges{} in the Supplementary Material. A one-way signed rank test indicates a statistically significant increase in the rankings of the government overthrow and civil war groups across all source countries, except for Germany, France, and the UK, at the 0.05 significance level.
We find that the ranks of Syria, Libya, Tunisia, and Jordan generally increased in the post-Arab Spring period. Egypt’s rank was already relatively high before the Arab Spring and remained relatively stable. This suggests that the shift in scholarly attention toward these target countries is also significant from the perspective of the source countries. Such a trend may reflect an academic response to regional developments and potential changes in funding priorities or broader shifts in research policy.

Because the distribution of foreign attention's contribution shares (\(A_i^*\)) follows a long-tailed pattern, making upward movement in the rankings is more challenging at lower ranks due to the inertia created by established research infrastructures. For example, a country advancing from rank 2 to rank 1 requires significantly more publications than a country moving from rank 147 to rank 146 among the 147 countries worldwide in our work. To account for these disparities, we measure the normalized rank change of country $i$ as \( NRC (A_i^*)= \frac{\Delta r(A_i^*)}{N - r(A_i^*)}\), which takes into account rank inertia \cite{iniguez2022dynamics}. Here, \(N\) represents the total number of countries (in this case, \(N = 147\)) and \(r(A_i^*)\) denotes the ranking of the country \(i\) within the global distribution of scholarly attention directed toward target countries. Figure \ref{geo_attention}b shows that, in relative terms, countries like Malaysia, Iran, and Qatar also saw a large change in their attention toward the target countries. Notably, Saudi Arabia and the UK appear in the top 5 countries that saw large relative as well as absolute changes.

When considering only Arab Spring-related research (as in Fig. \ref{arab_spring_related_research}), we observe that Western countries have also contributed significantly to scholarly attention (see \FigASAffCountries{} in the Supplementary Materials). However, Türkiye, Egypt, and Israel, countries within the region and directly or indirectly affected by the Arab Spring, also showed notable research activity. Meanwhile, Malaysia ranks 24th with 32 publications, while Saudi Arabia ranks 30th with only 23. A key reason for this may be that most Arab Spring-related research is concentrated in the social sciences, which is not the main focus of research originating from these two countries (\FigASSubjectAreas{} in Supplementary Materials).

Figures \ref{top_ten_attention_discipline}a–d show the top ten sources of foreign attention towards the region by computing the differences in yearly attention before and after the Arab Spring in the four main disciplines. Saudi Arabia emerges as a significant source of scholarly attention in the health, physical, and life sciences, with a particular focus on Egypt. Within the same disciplines, Iran's research focuses on Oman. Malaysia stands out as a key contributor to scholarly attention in the social and health sciences, particularly about Yemen. Spain's research focuses on Morocco and Tunisia, likely influenced by geographic proximity. In the social sciences, Western countries show a strong attention to Egypt and Syria, whereas their engagement in the life sciences is comparatively muted.

\subsection*{Difference-in-Differences estimation of the Arab Spring’s influence on scholarly attention}

To assess the impact of the Arab Spring on patterns of scholarly attention, we employ a difference-in-differences (DiD) statistical framework \cite{huntington2018chapter}. 
This approach, commonly used in econometrics and quantitative social sciences, involves employing observational data to simulate an experimental research design. It entails studying the differential effect of a treatment on a ``treatment group" versus a ``control group" in a natural experiment.
This approach allows us to estimate the average ``treatment '' effect of the Arab Spring by comparing changes over time in attention to our target countries, or in the three subgroups (government overthrow, civil war, and governmental changes) with those observed in three control groups: the rest of the world, a subset of MENA countries that remained relatively less affected during the same period, and comparable countries selected through coarsened exact matching based on pre–Arab Spring averages of GDP per capita, population size, and researcher population\cite{iacus2009cem}.
See the Materials and Methods section for a detailed explanation of the DiD approach and its application to our situation.

\begin{table}[!ht]
\centering
\resizebox{\textwidth}{!}{%
\begin{tabular}{l|c|c||
>{\columncolor{green!15}}c|
>{\columncolor{green!15}}c||
>{\columncolor{blue!15}}c
>{\columncolor{blue!15}}c
>{\columncolor{blue!15}}c|
>{\columncolor{blue!15}}c}
\toprule

\textbf{Dependent Variable} & \textbf{Control Variables} & \textbf{Control group} & \textbf{\(\beta_{\text{Target}}\)}& \(R^2 {\text{within}}\) & \textbf{\(\beta_{\text{GO}}\)} & \textbf{\(\beta_{\text{CW}}\)} & \textbf{\(\beta_{\text{GC}}\)} & \(R^2 {\text{within}}\)\\
\midrule
Log Attention               &\begin{tabular}{@{}c@{}} log GDP per capita, log pop.   \\log Rpop. \end{tabular}& Rest of world& \begin{tabular}{@{}c@{}}0.1226$^{**}$ \\ 
(0.049)\end{tabular}& 0.7034& \begin{tabular}{@{}c@{}}0.1226$^{***}$ \\ (0.049)\end{tabular} & \begin{tabular}{@{}c@{}}0.2549$^{***}$ \\ (0.096)\end{tabular} &\begin{tabular}{@{}c@{}}0.0195 \\ (0.052)\end{tabular}&0.7090\\
&  & Rest of MENA & \begin{tabular}{@{}c@{}}-0.0131\\ (0.067)\end{tabular}&0.6879& \begin{tabular}{@{}c@{}}0.1139 \\ (0.078)\end{tabular} & \begin{tabular}{@{}c@{}} 0.0811 \\ (0.131)\end{tabular} &\begin{tabular}{@{}c@{}}-0.1084$^{*}$ \\ (0.057)\end{tabular}&0.7091\\

&  & CEM group 1 & \begin{tabular}{@{}c@{}} 0.117$^{*}$\\ (0.067 )\end{tabular}&0.6686& \begin{tabular}{@{}c@{}}0.1578  $^{**,\dagger}$\\ (0.069)\end{tabular} & \begin{tabular}{@{}c@{}}  0.1338 \\ (0.095)\end{tabular} &\begin{tabular}{@{}c@{}}0.067 \\ (0.075)\end{tabular}&0.6687\\

\midrule
Log Domestic Attention               & log GDP per capita, log Rpop. & Rest of world& \begin{tabular}{@{}c@{}}-0.0901 \\ (0.066)\end{tabular}&0.8053 & \begin{tabular}{@{}c@{}}0.1571 $^{***,\dagger }$ \\ (0.045)\end{tabular} & \begin{tabular}{@{}c@{}}-0.2896$^{*** }$ \\ (0.027)\end{tabular} &\begin{tabular}{@{}c@{}}-0.1128 \\ (0.074)\end{tabular}&0.8042\\
&  & Rest of MENA & \begin{tabular}{@{}c@{}} -0.0029 \\ (0.106)\end{tabular}&0.8807 & \begin{tabular}{@{}c@{}} 0.2451$^{*** }$ \\ (0.087)\end{tabular} & \begin{tabular}{@{}c@{}}-0.2213$^{**}$ \\ (0.093)\end{tabular} &\begin{tabular}{@{}c@{}}-0.0208 \\ (0.103)\end{tabular}&0.8838\\

&  & CEM group 2 & \begin{tabular}{@{}c@{}} -0.1912\\ (0.1307)\end{tabular}&0.8038 & \begin{tabular}{@{}c@{}}  0.0383  \\ (0.1101)\end{tabular} & \begin{tabular}{@{}c@{}}-0.3898$^{***,\dagger }$ \\ (0.1264)\end{tabular} &\begin{tabular}{@{}c@{}}-0.2145 $^{* }$ \\ (0.1306)\end{tabular}&0.8015\\
\midrule

Log Foreign Attention               & log GDP per capita, log pop.   & Rest of world&\begin{tabular}{@{}c@{}}0.1355 \\ ( 0.106)\end{tabular}&0.2767 & \begin{tabular}{@{}c@{}}0.1462 $^{**}$ \\ (0.071)\end{tabular} & \begin{tabular}{@{}c@{}}0.5770$^{***}$ \\ ( 0.206)\end{tabular} &\begin{tabular}{@{}c@{}}-0.1258 $^{**}$ \\ ( 0.057)\end{tabular}&0.3367\\
& & Rest of MENA &\begin{tabular}{@{}c@{}} -0.1209 \\ ( 0.109 )\end{tabular}& -0.1439 & \begin{tabular}{@{}c@{}} -0.0270 \\ (0.087)\end{tabular} & \begin{tabular}{@{}c@{}} 0.2165 \\ (0.190)\end{tabular} &\begin{tabular}{@{}c@{}}-0.2663 $^{***}$ \\ ( 0.096)\end{tabular}&0.0776\\

& & CEM group 3 &\begin{tabular}{@{}c@{}} 0.0126\\ ( 0.0107)\end{tabular}& -0.0091 & \begin{tabular}{@{}c@{}}  0.1514$^{***, \dagger}$ \\ (0.043)\end{tabular} & \begin{tabular}{@{}c@{}}  0.3987 $^{***, \dagger}$\\ ( 0.149)\end{tabular} &\begin{tabular}{@{}c@{}}-0.0504 \\ ( 0.0691)\end{tabular}& 0.1587\\
\bottomrule
\end{tabular}%
}
\caption{\textbf{Difference in differences estimates of the effect of Arab Spring on scholarly attention.} Estimated coefficients from difference-in-differences regressions evaluating the post-treatment effects of the Arab Spring on log attention (total, domestic, foreign). Models control for log GDP, population, and research population, where applicable. Each colored column represents estimates from separate regressions, comparing the target countries with three control groups: the rest of the world and MENA countries less affected by the Arab Spring and countries found through coarsened exact matching. Standard errors are in parentheses. ***, **, and * denote statistical significance at the 1\%, 5\%, and 10\% levels, respectively. $\dagger$ denotes regressions where the parallel prior trend test fails to reject the null hypothesis of no difference in pre-Arab Spring trends at a 5\% significance level.}
\label{tab:attention_DiD}
\end{table}
Table \ref{tab:attention_DiD} presents the difference-in-differences (DiD) analysis of scholarly attention. The columns labeled \(\beta\) report the estimated effect of the 2011 Arab Spring on scholarly attention to the target countries.
The analysis reveals that the Arab Spring had a significant positive effect ($p$ $<$ 0.05) on scholarly attention towards affected countries compared to the rest of the world, and controlling for GDP, population, and researchers' population. Similar results ($ p < 0.1$) also hold when comparing to similar countries identified by coarsened exact matching (CEM). 

At the subgroup level, countries experiencing government overthrows and civil wars exhibit a positive treatment effect on total attention, with the civil war subgroup experiencing a stronger effect. In contrast, the effect for countries undergoing governmental changes is only statistically significant when compared to the rest of the MENA region, and it is negative. When using a CEM-based control group, the estimates for government overthrow and civil war remain positive; however, the validity of these comparisons is limited by the failure to satisfy the parallel trends assumption ($p < 0.05$) or test power. Similarly, the negative effect of governmental changes observed in the MENA comparison is not replicated in the CEM specification, further suggesting that results are sensitive to the choice of control group.

Further disaggregation by the source of attention offers additional insights. The effect on domestic attention is negative in the civil war subgroup compared to the rest of the world, the MENA region. In contrast, the effect is positive in the government overthrow subgroup, possibly reflecting the role of reform strategies. When using a CEM-based control group, the negative effect of civil wars remains, but the estimates again fail the parallel trends test, limiting their interpretability. For governmental changes, the CEM comparison replicates the negative effect observed in the broader control groups, this time significant at ($p < 0.1$).

Regarding foreign attention, the effects are positive and statistically significant in the government overthrow and civil war subgroups, with the civil war subgroup exhibiting a notably higher effect. Conversely, the effect on foreign attention is negative in the governmental changes subgroup compared to the rest of the world and the MENA region. When the DiD analysis is conducted using the CEM control group, the government overthrow and civil war subgroups still show positive and statistically significant effects; however, because the estimates fail the parallel trends test, we refrain from drawing substantive conclusions from them.

Collectively, these findings provide statistical evidence that the Arab Spring had an impact on scholarly attention toward the target countries, underscoring the heterogeneity of this impact across the subgroups. While political instability, via government overthrows and civil wars, has sparked increased academic interest, governmental changes may dampen this engagement.

\subsection*{Interplay between scholarly attention and funding}

\begin{figure}[!ht]
    \centering
    \includegraphics[width=\linewidth]{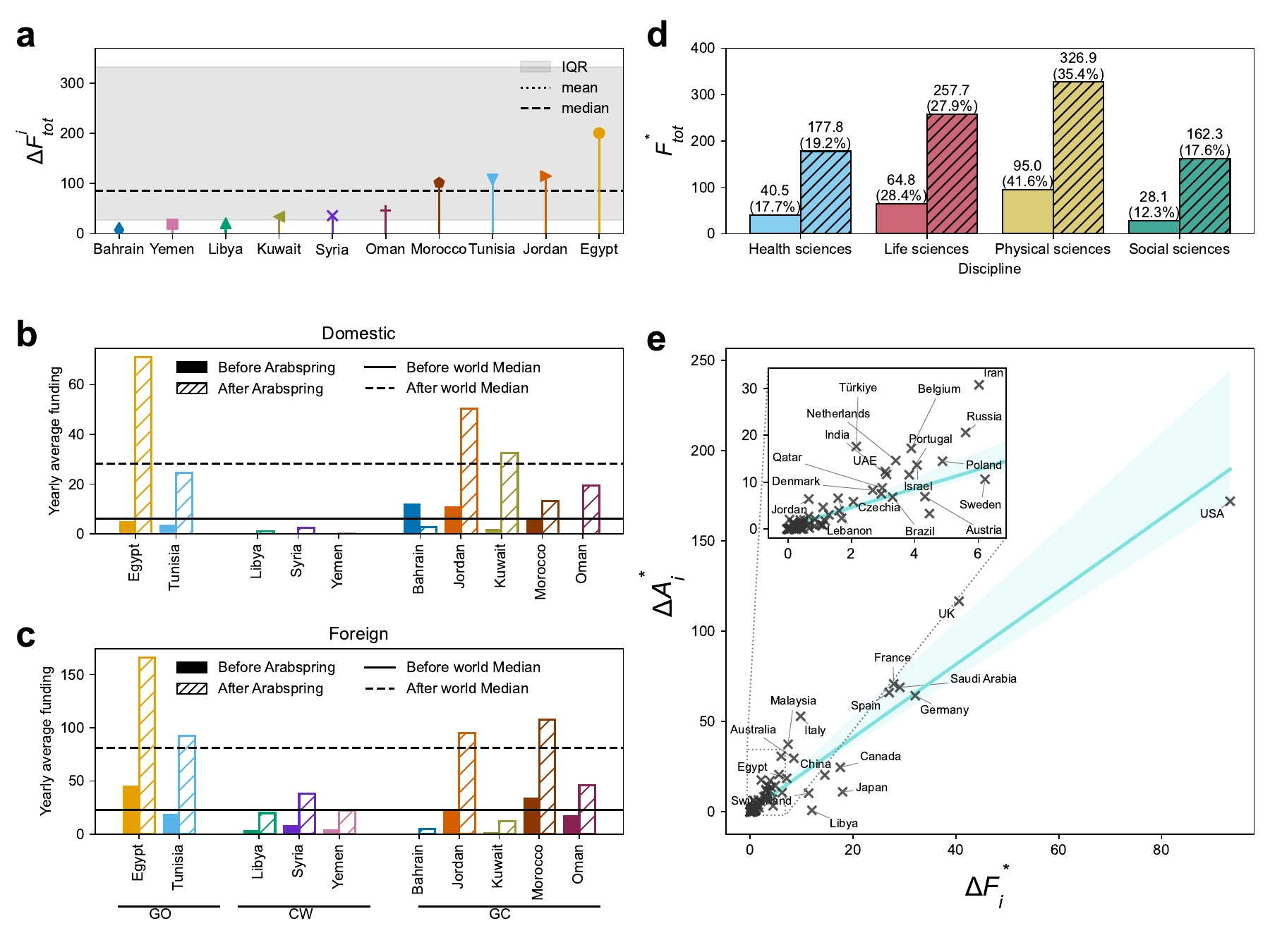}
    \caption{\textbf{Shifts in funding patterns before and after the Arab Spring.} (\textbf{a}) Changes in funding before and after the Arab Spring per target country. The dashed line, dotted line, and shaded gray area represent the world's mean, median, and interquartile range.
    We observe substantial increases for Egypt, Tunisia, Jordan, and Morocco.
    (\textbf{b}) \& (\textbf{c}) Yearly average domestic and foreign funding before and after the Arab Spring per target country.
    Domestic funding surged for Egypt and the governmental changes (GC) group, particularly Jordan and Kuwait. Foreign funding grows for the government overthrow group (GO), Morocco, and Jordan.
    (\textbf{d}) Total funding linked with scholarly attention toward the target countries per discipline. Plain and hashed bars indicate values before and after the Arab Spring, respectively.
    Although the social sciences hold the smallest share of total funding, their allocation shows the largest increase from 12.4\%  to 17.6\%. (\textbf{e}) Change in scholarly attention towards the target countries as a function of changes in funding per source country.}
    \label{funding}
\end{figure}

A way for countries to shift scholarly attention is through their allocation of research funding. For each paper in our dataset, we extract the countries providing funding from the article's acknowledgment section.
Similarly to scholarly attention, we model the funding patterns with a directed weighted network. Nodes once more represent countries, and directed edges capture the funding allocated from country \(i\) to study country \(j\), namely \(F^j_i\) (see Materials and Methods for details). In this way, we measure funding linked to scholarly attention towards the target countries as the number of papers published per year that mention the target countries and have received funding from specific countries.

Figure \ref{funding}a displays the changes in funding, $\Delta F_i^*$, received before and after the Arab Spring. The rise in funding for Egypt, Tunisia, Jordan, and Morocco exceeds the median. Domestic funding for studying the government overthrow and governmental changes groups, except Bahrain, has greatly increased (Fig. \ref{funding}b). The changes for Egypt, Tunisia, and Kuwait are notable, from below the median to above the median. Meanwhile, foreign funding for research on all target countries has grown substantially (Fig. \ref{funding}c). 

Similarly to the patterns observed in scholarly attention, total funding for research directed at these target countries ($F^*_{tot}$) increased across all disciplines following the Arab Spring (Fig. \ref{funding}d). Although the social sciences received the smallest overall share of funding in both periods, their proportion increased from 12.4\% to 17.6\%. This growth may explain the rise in scholarly attention to the social sciences, which likely aims to assist researchers in interpreting causes, consequences, and long-term societal impacts.

Funding and scholarly attention towards target countries exhibit a strong correlation (Fig. \ref{funding}e), with a Pearson correlation coefficient of 0.94 and a Spearman rank correlation of 0.90 (\(p < 0.0001\)). The United States leads in funding changes and scholarly attention, followed by Saudi Arabia, France, Spain, and Germany. This ranking suggests that countries with stronger economies tend to have more available funding. However, the Pearson and Spearman correlation coefficients between funding and GDP per capita are only 0.383 and 0.556, respectively (\(p < 0.001\)), indicating only a moderate correlation (see \FigFundingVsGDP{} in Supplementary Materials). Subsequently, while GDP plays a role in funding availability, some countries provide disproportionately higher funding than others. Countries like Iran, Malaysia, and Australia, despite making minor changes in funding, made relatively large changes in their scholarly attention to the target countries. 

To assess whether the Arab Spring significantly influenced funding patterns, we applied the same difference-in-differences framework used for scholarly attention (see Materials and Methods), using the log of total, domestic, and foreign funding as outcome variables while controlling for the log of population and GDP per capita. The results, presented in \TabDiDFunding{} of the Supplementary Materials, indicate that, for some regressions, the pre-treatment parallel trends assumption is not satisfied, and the model fit is generally poor. Nonetheless, the remaining regressions that still meet the required assumptions yield results that align with the patterns shown in Figure \ref{funding} (see Table \TabDiDFunding{} in the Supplementary Materials). The Civil War group shows a significant post-Arab Spring positive effect on funding relative to the rest of the world ( $p< 0.01$). The effect is also positive on domestic funding for the governmental overthrow group and foreign funding for the civil war group (both significant at $p<0.01$ ). Together, these results underscore a post-uprising shift in both the level and source of funding, varying by political trajectory.

\subsection*{Interplay between scholarly attention and scholarly migration}
Significant social or political phenomena, such as a revolution or civil war, typically affect a country's mobility of scholars. These events can create a brain drain, where the number of researchers leaving the country exceeds those entering, or a brain gain, where researchers who had previously left due to political oppression and dissatisfaction with the former government may return as conditions improve. In the former, the net migration rate of researchers increases, and in the latter, it decreases.

In the context of increased scholarly attention, two scenarios can be considered. First, scientists who leave their country due to political upheavals may remain interested in their homeland and continue conducting research related to it. Second, scientists who return to their country after positive political and economic changes may be motivated to contribute to their country's advancement. This leads to increased domestic knowledge through their intensified research efforts. With this background, we examine whether rising scholarly attention from a given country to the target countries is accompanied by an increase in scholarly migration to that country.
\begin{figure}[!ht]
    \centering
    \includegraphics[width=\linewidth]{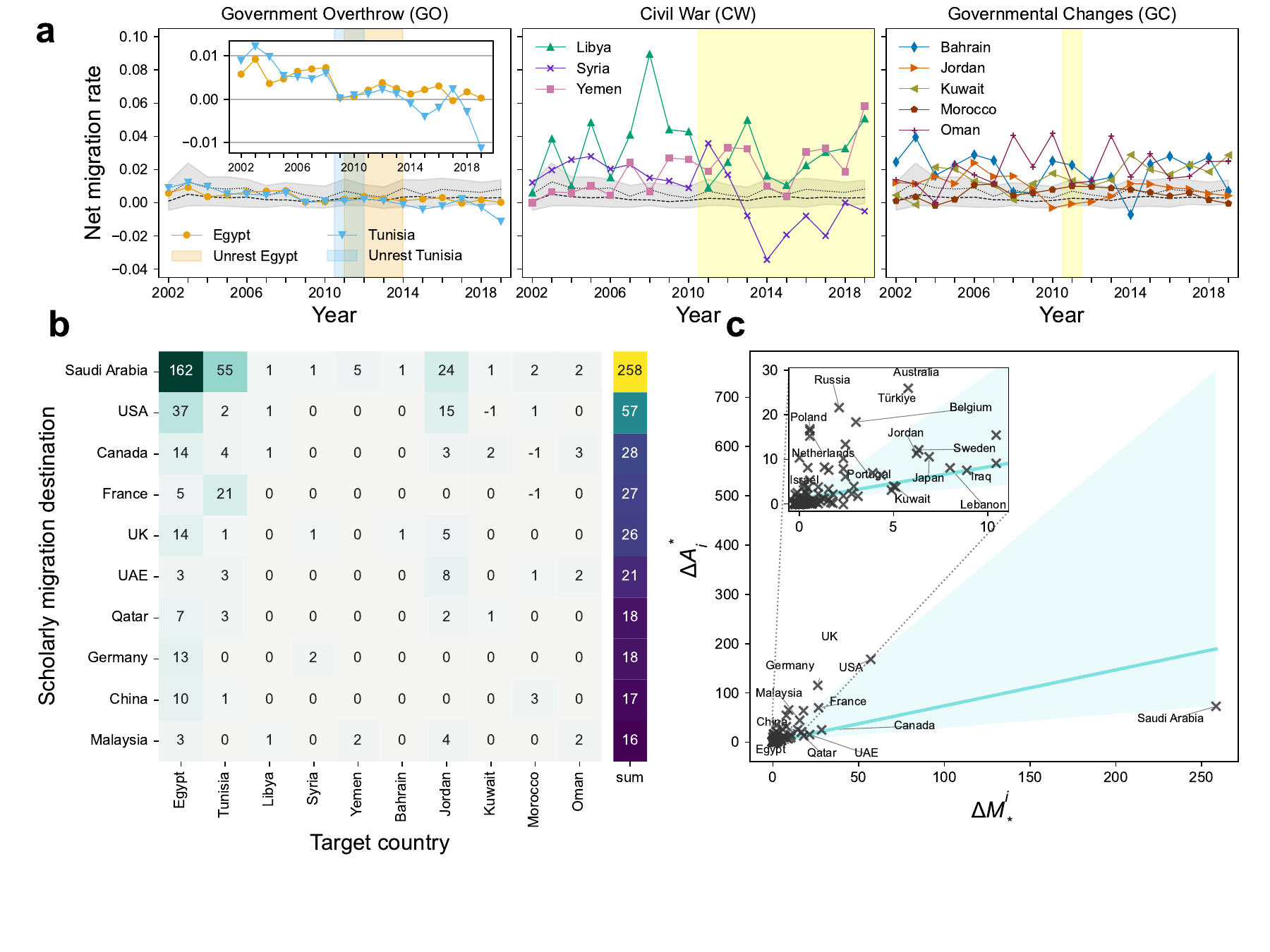}
    \caption{ \textbf{ Shifts in migration patterns before and after the Arab Spring} (\textbf{a}), Net migration of researchers for target countries before and after the Arab Spring. In all panels, the gray area represents the interquartile range for global trends, with the dashed line indicating the median and the dotted line showing the mean. The inset in panel a (GO) shows the relevant values for Egypt and Tunisia. (\textbf{b}), Changes in the number of migrations from target countries to destination countries before and after the Arab Spring. We show the top 10 destinations.
    Islamic and Western countries are among the top 10 migration destinations for researchers from target countries, with special bonds between Saudi Arabia and Egypt. (\textbf{c}) Change in scholarly attention towards the target countries as a function of changes in migrations from the target countries.}
    \label{migration}
\end{figure}

We track scholarly migration from and to the target countries using the scholarly migration database created using changes in affiliations recorded in Scopus-indexed articles \cite{akbaritabar2024bilateral, ScholarlyMigrationDatabase2022}.
Figure \ref{migration} shows the shifts in migration patterns before and after the Arab Spring.
Before the Arab Spring, Egypt consistently attracted researchers, evidenced by positive net migration rates, except in 2008, when the rate nearly reached zero, which could be attributed to unfavorable economic conditions (Fig. \ref{migration}a). With the onset of the Arab Spring, an increase in the net migration rate of researchers is observed. This may be due to two factors: researchers who had previously left the country due to economic inefficiencies and political turmoil might have returned because of improved conditions. Additionally, Egypt, being a relatively stable country in the region with reputable universities, has attracted researchers from other target countries. However, after 2015, the net migration rate tends to decrease, potentially due to the effects of the Arab Winter, the resurgence of authoritarianism and Islamic extremism in some target countries. Tunisia exhibits a downward trend, transitioning from a country that attracted talent to one experiencing a brain drain. This process began in 2011, and the only subsequent year when the net migration rate was not negative was 2016, a significant year in Tunisia's political landscape. In 2016, Tunisia experienced widespread protests triggered by high unemployment, particularly among young graduates, and Tunisia's prominent political party underwent a significant shift. Syria began experiencing a brain drain immediately following the Arab Spring. Meanwhile, patterns for Yemen and Libya are less informative due to their relatively small number of researchers compared to other countries. No clear pattern is observed within the group of countries undergoing governmental changes.

We now model scholarly migration as a directed network, \( M \), where nodes represent countries and directed edges, \( M_{i}^j \), indicate the number of scholars migrating from country \( i \) to country \( j \). Figure \ref{migration}b shows the ten destination countries that saw the largest increase in migration of researchers from the target countries. The analysis reveals that Saudi Arabia, Qatar, and the UAE are among the top ten destinations, which could be attributed to their robust economies, geographical proximity to the target countries, and shared language and religion \cite{el-ouahi_analyzing_2021}. Post Arab Spring, 162 more scientists affiliated with Egyptian institutes have relocated to Saudi Arabia each year. Malaysia, a predominantly Islamic country, emerges as a growing destination for researchers. Researchers from the target countries migrated to Islamic countries at significantly higher rates compared to countries with other religions, both before and after the Arab Spring. While 5.1\% of worldwide immigrant scholars migrated to Islamic countries before and 8.5\% after the Arab Spring, the corresponding values for migrant scholars from the target countries were substantially higher, at 33.5\% before and 46.2\% after (see \FigReligionMigration{} in Supplementary Materials). 

Western countries, such as the UK, USA, France, and Germany, are also in the top 10 destinations, likely due to their strong economies and leadership in scientific research. The 10 destination countries are classified as HIC (high-income countries) and Upper-Middle-Income Countries (UMIC) in the Worldbank classification.  There exists a weak, significant correlation between the change of migration to a country and its average GDP in the period (Spearman Correlation Coefficient: 0.33, \(p < 0.0001\), see \FigMigrationVsGDP{} in Supplementary Materials). 

A positive correlation exists between changes in scholarly migration from target countries, \(\Delta M^i_*\), and change in scholarly attention towards them, \(\Delta A_i^*\), (Fig. \ref{migration}c), demonstrated by a Pearson correlation coefficient of 0.505 and a Spearman rank correlation of 0.725, both of which are highly significant (\(p < 0.0001\)). This alignment suggests that migration patterns may partly explain shifts in scholarly attention.

\section*{Discussion}
Our findings reveal that scholarly attention toward the ten countries affected by the Arab Spring has significantly increased in the post-Arab Spring period compared to the rest of the world, as measured by a difference-in-differences approach, and that this increase aligns with changes in funding and scholarly migration. 

Changes in scholarly attention cannot be attributed solely to the Arab Spring, as broader global or regional political, economic, and social dynamics likely contributed to shaping research trends. For instance, the long-term effects of the 2008 economic downturn, as well as the Arab Winter \cite{King_2020} (a term used to describe the resurgence of authoritarianism, political violence, and civil conflict in the aftermath of the Arab Spring uprisings) and the rise of ISIS in Syria, have also influenced science. Topic analysis revealed themes such as social media, oil markets, women's rights, refugees, and regional geopolitics, illustrating how different political trajectories shaped the evolution of research agendas of studying the Arab Spring itself. 

We find varied patterns of foreign-to-domestic attention ratio across the target countries. For example, Egypt and Tunisia received more domestic attention than foreign attention before and after the Arab Spring. Countries experiencing civil wars received more foreign attention before the Arab Spring, and this imbalance has grown even more pronounced post-Arab Spring, particularly in the social sciences and health sciences. Our findings show that shifts in scholarly attention are correlated with changes in research funding. Funding plays a crucial role in setting research agendas and influencing policy directions. While domestic funding has increased in countries experiencing government overthrows or governmental changes, foreign funding of research about the target countries has seen a more pronounced rise, particularly in nations affected by civil wars. Notably, the increase in funding has been most significant in the social sciences. 

These findings raise important questions about the optimal balance between domestic and foreign scholarly attention and its associated funding in fostering a sustainable research discourse. Domestic research and funding are crucial in strengthening national scientific productivity and research ecosystems, allowing local scholars to address context-specific challenges and build long-term research capacity. Meanwhile, foreign attention and funding enhance visibility and productivity and may facilitate integration into global academic networks. However, an overreliance on foreign attention may marginalize local voices in shaping discourse, potentially shifting knowledge production toward external and often temporary agendas rather than aligning with national priorities. Moreover, the temporary nature of such foreign engagements makes them susceptible to being overshadowed by events elsewhere or newer trends.

Our findings also highlight the varied dynamics of scholarly migration: Egypt exhibited positive net migration rates, but Tunisia and Syria faced significant brain drain. The interplay between scholarly migration and scholarly attention may partly reflect institutional hiring practices, whereby researchers are employed due to their expertise in their country of origin \cite{kohstall_academia_2018}. In other cases, researchers may intentionally focus their work on their home country as an expression of national commitment, particularly in times of crisis, motivated by the hope of contributing to long-term development \cite{dajani_scientists_2023}. However, the mechanisms linking scholarly migration and patterns of scholarly attention remain underexplored, and further qualitative and quantitative research is needed to understand these relationships better.

Our analysis reveals the prominence of Saudi Arabia as a source of scholarly attention and funding, as well as a destination for academic migration. This may be attributed to its increased research productivity and impact \cite{turki_discussing_2019}. Government initiatives have primarily driven this growth since the late 2000s, with the aim of transforming the country's major universities into world-class institutions. Saudi Arabia has been reported as one of the countries experiencing the highest surge in hyperprolific researchers—those publishing more than 72 papers annually. The role of cash incentives \cite{abritis2017cash} in this trend has raised concerns about the potential use of questionable methods to boost publication counts \cite{conroy2024surge}. However, our analysis indicates that these researchers had minimal impact on our results (see ``Analysis of Hyperprolific authors''  in the Supplementary Materials).

Our study has two main limitations. First, it relies on Scopus-indexed publications, which primarily index English-language journals. As a result, contributions from local researchers publishing in non-English outlets may be underrepresented. Although accessing comprehensive local publication databases would offer valuable insight, doing so would require a multilingual country-detection algorithm and a scalable processing pipeline, both of which fall beyond the scope of this study. Furthermore, the quality, accessibility, and consistency of indexing in many local journals remain uneven, making it difficult to assess their reliability and integrate them systematically. Future research should address these gaps by developing tools to incorporate non-English and regionally published scholarship more effectively.
Second, we measured scholarly attention by extracting mentions of countries in research articles. However, the absence of a country's mention in a research paper does not necessarily indicate a lack of focus on that country, just as the presence of a country's name does not automatically mean the study is centered on it. 

Despite these limitations, our study is the first large-scale investigation of the Arab Spring's impact on scholarly attention and its relationship with funding and scholarly migration. Our framework enabled us to identify the significantly different trajectories taken by countries affected by the Arab Spring, as well as the foreign countries that emerged as key players in shaping academic research on the region.

This framework, comprising three lenses---scholarly attention, research funding, and scholarly migration---offers a scalable approach for constructing a comprehensive scientometric portrait of regional and global events. By combining it with empirical methods, such as quasi-experimental methods, it has wide applicability in the science of science \cite{liu2023data}.

For example, it can be applied to understand the impact of other conflicts, such as the Russia–Ukraine war, enabling a systematic investigation of how research priorities were reshaped: where scientific attention was directed, who funded the response, and how patterns of scholarly migration evolved. The framework can also be employed to assess the impact of political transitions, particularly changes in leadership or governance models, on both the international scholarly attention a country receives and the attention it directs outward. In this way, the framework provides a novel lens for observing how science policy agendas emerge, evolve, and circulate within the global research landscape.

\section*{Materials and Methods}

\subsection*{Publication dataset}
This study analyzes over 25 million journal articles indexed in Scopus, excluding other document types such as books, editorials, conference papers, and review letters. Of these, 3.7 million articles (14.4\%) mention at least one country, representing a substantial and still underexplored source of information on the geographic dimensions of scientific research. From this subset, we focus on approximately 73 thousand articles ($\approx$ 2\%) that reference at least one of the countries directly affected by the Arab Spring: Bahrain, Egypt, Jordan, Kuwait, Libya, Morocco, Oman, Syria, Tunisia, and Yemen.

The temporal scope of our analysis spans from 2002 to 2019, covering the nine years preceding and following the onset of the Arab Spring in 2011, to capture the scholarly landscape surrounding this sociopolitical event. We chose not to analyze papers published after 2019 to avoid the disruptions caused by the COVID-19 pandemic, which significantly altered patterns of scientific mobility, collaboration, and scholarly attention. We focus on articles with English titles and abstracts to simplify the analysis of country mentions, avoiding the challenges of multilingual variations in country name spellings.

We extract four key dimensions of information from each publication. The first dimension is the subject area, determined using the Scopus ASJC classification scheme, which categorizes publications into 27 subject areas across four major areas (see Supplementary Materials, Table \ref{tab: scopus} for the classification). Disciplines were counted in a manner that acknowledges the interdisciplinary nature of many research papers. If a paper is classified under multiple disciplines, such as Discipline A and B, it is counted as half a contribution to each discipline. This method ensures a more accurate distribution of research efforts across disciplines.

The second dimension is the countries of affiliation of the authors, inferred from the authors' institutional affiliations as provided by Scopus. The third dimension is the funding countries, identified from the acknowledgment sections, where Scopus captures information about funding sources. Funding sources at a continental or global scale, such as the European Union's Horizon program, are identified in the Scopus dataset but excluded from this analysis, as our focus is on country-level dynamics. The fourth dimension refers to the mentioned country, which is explicitly mentioned in the publication and identified from the titles and abstracts\cite{castro2022north}. We interpret these mentions as a form of scholarly attention to countries, reflecting the research's focus and relevance to specific geographical contexts. The methodology for identifying the countries mentioned is discussed in the next section.

\subsection*{Identifying scholarly attention to countries with country name extraction}
To extract mentions of countries or regions, we followed several steps. First, we removed copyright text from abstracts by deleting any content appearing after terms like '@' or 'Copyright (C)', as such text may include the publisher's country, which does not reflect the geographical focus of the research. Next, we used the Geotext library \cite{Software_geotext2024} to identify and extract country names from titles and abstracts. Geotext employs country names, variations, and abbreviations from GeoNames, a comprehensive geographical database, to detect and classify mentions of cities and countries in textual data. Additionally, we used the pycountry library \cite{Software_pycountry2024} to map the extracted country names to their corresponding ISO Alpha-3 country codes, ensuring standardized representation. Then, we addressed specific cases to improve accuracy. Terms such as `US dollar,' `New Mexico,' `USD,' `HK,' and `Congo Red,' and `Michael Jordan, ' which contain geographical names in text but do not necessarily indicate the geographical focus of the research, were excluded from the list. These steps ensured that the extracted mentions accurately reflect scholarly attention to countries or regions.

We also excluded demonyms (e.g., ``Syrian,” ``Egyptian”), as their presence does not necessarily indicate that the study focuses on the corresponding country. For example, research on the Syrian population in Lebanon or Germany may primarily address migration or diaspora dynamics, rather than Syria itself. In addition, many demonyms overlap with language names (e.g., French, Vietnamese, Georgian), creating further ambiguity, as such terms may refer to the language rather than the people or the country. Conversely, when research is conducted within a country, such as a study on ``Egyptian goats” based in Egypt, the country is typically mentioned explicitly in the abstract or title, increasing the likelihood of accurate identification with only direct country name matching.

To enhance the algorithm's accuracy, variations in country name spellings were incorporated into the dictionary and matched with regular expressions. For example, for the USA, `U.S.,' `United States,' `United-States,' `United States of America,' `US', and `U.S.A.' were added to the dictionary.

We applied post-processing using specific criteria to enhance the accuracy of country identification. First, we excluded countries with populations under 1 million as of 2019, focusing our analysis on more populous nations. Second, we encountered instances where country names could lead to false positives due to their similarity with terms commonly used in academic disciplines. For example, "Jordan" shares its name with concepts in physics, mathematics, computer science, and engineering, named after Camille Jordan and Pascual Jordan (Including, but not limited to, Jordan Algebra, Jordan Normal Form, Jordan Measure, Jordan Matrix, and the Jordan–Wigner Transformation). We avoided including these subject areas to mitigate inaccuracies when studying the target countries. Third, we chose to exclude physics for another reason: the prevalence of large-scale (Mega) collaborations in this field, often involving hundreds of authors from multiple institutions and countries, makes it challenging to assess individual contributions to a single paper accurately \cite{rossi2019bibliometrics}.

Additionally, geopolitical sensitivities required additional adjustments. Cyprus and Macedonia were excluded because of their complex political divisions. Likewise, countries with "Guinea" in their names, Guinea, Guinea-Bissau, Equatorial Guinea, and Papua New Guinea, were removed due to difficulties differentiating them using dictionary-based methods and their shared name with guinea pigs, a frequent subject in scientific and laboratory contexts. Similar exclusions were made for Ireland and Northern Ireland, the Republic of the Congo and the Democratic Republic of the Congo, as well as Sudan and South Sudan. Countries that gained independence during our research period were also removed to avoid complications. Kosovo declared independence from Serbia in 2008, though its recognition remains disputed. Montenegro and Serbia became independent entities following the dissolution of the State Union of Serbia and Montenegro in 2006. East Timor (Timor-Leste) achieved full independence from Indonesia in 2002 after a UN-led transition. 

\subsection*{Assessing the accuracy of the country extraction algorithms}
We evaluate the accuracy of the identification process through three tasks: first, assessing whether our method can detect country mentions; second, determining how accurate and comprehensive the method is in identifying these mentions; and third, evaluating the accuracy of detecting mentions of the focal country in this study. To achieve this, we created three samples of articles, totaling 1,000 articles, with titles and abstracts obtained for analysis. The first sample comprises 520 randomly selected papers from our analyzed time frame, with 20 papers chosen from each of the 27 subject areas, including those with and without country mentions. The second sample comprises 280 papers that mention countries without restrictions on subject areas. The third sample includes 200 randomly selected papers that mention at least one of the 10 target countries analyzed in this study, with 20 documents mentioning each country.

Each paper was independently annotated by two coders, who are among the authors of this paper. They identified all countries or regions mentioned in the titles or abstracts. Coders were instructed not to annotate instances where country names appeared but did not refer to geographical locations, following the algorithm's criteria. The inter-coder agreement rate was 89.1\%, reflecting a high level of consistency in identifying geographical references. Then, two baselines are implemented: one derived from the intersection of the sets of countries found by the two annotators and the other from their union. 
We use two key measures commonly used in this domain: Accuracy/ Exact Match Ratio \cite{sokolova2009systematic} and Average Jaccard Similarity.

The Accuracy or Exact Match Ratio \cite{sokolova2009systematic}   evaluates the proportion of instances where the model's predicted labels exactly match the true labels and is given by
\[
\text{Accuracy} = \frac{\sum_{i=1}^{n} I(L_i^d = L_i^c)}{n},
\]
Where \(n\) is the total number of instances, \(L_i^d\) represents the data labels for the text \(i\), \(L_i^c\) represents the labels predicted by the model, for instance, \(i\), and \(I\) is the indicator function that equals 1 if \(L_i^d = L_i^c\) and 0 otherwise. A high value in this metric indicates consistent, perfect predictions for all labels in each instance.
This metric is highly strict, requiring a perfect match between predicted and actual label sets for each instance.

The Average Jaccard similarity provides a more flexible evaluation by considering the partial overlap between the model’s output and the baseline, calculating the ratio of shared elements to the total unique elements across both sets (intersection over union). It is given by
\[
\text{Average Jaccard Coefficient} = \frac{1}{n}\sum_{i=1}^{n} \dfrac{ L_i^c \cap L_i^d}{L_i^c \cup L_i^d}.
\]
For example, if the model predicts entities \(\mathcal{A}=\left\{ A, B\right\} \) and the human annotation is \(\mathcal{B}=\left\{B, C\right\}\), the Jaccard score is \(\frac{|\mathcal{A}\cap\mathcal{B}|}{|\mathcal{A}\cup\mathcal{B}|}=\frac{1}{3}\), reflecting partial agreement, while accuracy is 0.

Our identification algorithm is evaluated against three established baselines: specialized transformer models such as BERT-base NER and BERT-large NER \cite{DBLP:journals/corr/abs-1810-04805}, as well as the large language model ChatGPT-4.0 \cite{chatgpt4o2025}.

For ChatGPT, we used the following prompt with a temperature of 0.3 (for high precision):
\begin{tcolorbox}[colback=white, colframe=white, boxrule=1pt, sharp corners]
\begin{adjustbox}{minipage=\linewidth,scale=0.9, center}
As a named entity recognition tool, you are responsible for analyzing academic texts from various sources. Your task is to find all the country names mentioned in the text.
Constraint: Answer with only the Python list of countries in ISO 3 format mentioned in the text that is most accurate, and nothing else.
Text:{text}
\end{adjustbox}
\end{tcolorbox}

As shown in Table \ref{tab:performance_metrics} in the Supplementary Materials, our algorithm outperforms all other models, except for those mentioning target countries, where ChatGPT\cite{chatgpt4o2025} slightly outperforms it. In this task, our dictionary-based approach offers superior general accuracy and is significantly more efficient in terms of computation speed and cost. Running a large language model, such as ChatGPT, on millions of articles would currently cost tens of thousands of dollars and take anywhere from days to weeks, depending on the implementation.

\subsection*{Network of scholarly attention}

\begin{figure}[!ht]
    \centering
    \includegraphics[width=\linewidth]{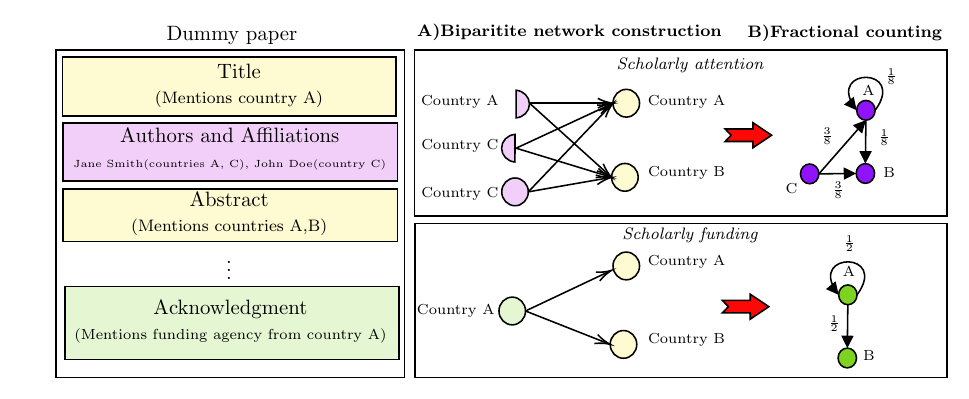}
\caption{\textbf{Construction of the scholarly attention and funding networks}. We first extract information from each paper, including the list of authors and their country of affiliation, and identify countries mentioned in the title and abstract. Additionally, the countries of the funding agencies are retrieved if provided. We then proceed in two steps: (A) Construction of a bipartite network, with one set of nodes representing authorship/funding countries and the other representing the mentioned countries. (B) Projection of the bipartite network to a unipartite weighted directed network using fractional counting to ensure that each paper contributes equally to the network.}
    \label{fig0}
\end{figure}

Figure \ref{fig0} illustrates the construction of the scholarly attention network, which follows a two-step procedure after identifying the country names in each paper:  Bipartite network construction and network projection using fractional counting.

We first construct a bipartite network that connects two distinct sets of nodes: on one side, the countries corresponding to the authors' affiliations, and on the other, the countries mentioned in the publication's title or abstract. 

Since many publications have multiple authors with multiple affiliations, we employ fractional counting to ensure that each author's contribution is accurately represented. For example, consider a dummy paper authored by Jane Smith and John Doe. Suppose Jane Smith is affiliated with Country A and Country C, meaning her contribution is split equally (0.5 for each country). In contrast, John Doe, who is only affiliated with Country C, contributes a full count. 

If the publication’s title or abstract mentions Countries A and B, these mentions are linked to the corresponding affiliation nodes in the bipartite network. In this way, each publication generates a set of edges that connect affiliation countries to the mentioned countries, reflecting the flow of scholarly attention from the research origin to the research subject.

Once the bipartite network is established, it is projected onto a unipartite, directed network. In this projection, the nodes remain as countries, but the edges are now directed from the affiliation countries (sources) to the mentioned countries (targets). The weight of an edge from country \(j\) to country \(i\), denoted by \(A^i_j\), reflects the fractionally counted number of articles in which country \(i\) is mentioned by authors affiliated with institutions in country \(j\).

Fractional counting is crucial in this process because it normalizes the contributions when multiple affiliations or mentions occur. For each publication, the total weight across all outgoing edges is normalized so that it sums to one. In contrast, using full counting, where edge weights vary based on the number of authors or countries, increases the risk of misunderstanding or misinterpretation \cite{perianes2016constructing}.

\subsection*{Network of scholarly funding}
ICSR lab extracts funding information, when present, from the funding or acknowledgment sections of research papers. 
The funding information includes the country of origin for the founding agencies. We use this information to construct the scholarly funding network similarly to the scholarly attention network (see Fig. \ref{fig0}).
This process involves two steps: constructing a bipartite network and subsequently projecting it into a unipartite network using fractional counting.

In the bipartite network construction, two distinct sets of nodes are defined. On one side are the countries of the funding agencies, and on the other are those mentioned in the publication's title or abstract. The bipartite network is converted into a unipartite, directed network using fractional counting. In this structure, countries serve as nodes, and the edges, \(F^i_j\), represent the fractionally counted number of publications that mention country \(i\) and are funded by agencies from country \(j\).

\subsection*{Arab Spring-related publications}
In Figure \ref{arab_spring_related_research}, we conducted an analysis to assess the extent to which the surge in scholarly attention received in target countries is attributed to research related to the Arab Spring. Scopus \cite{scopus2025} is a bibliometric database that allows users to filter publications by submitting information through a `query.'

The query was used to identify academic literature on the Arab Spring in Scopus-indexed sources. It did not include specific terms associated with the countries impacted by the Arab Spring, as our objective was to capture the overarching trend. Additionally, we included more general terms, such as "civil unrest" and "protest," which resulted in the retrieval of some articles published before 2011. Our decision was based on two considerations. First, it is challenging to distinguish between publications explicitly focused on the Arab Spring and those that address other social events in the region, such as earlier revolutions or uprisings. Second, we believed employing these broader terms would effectively highlight the increasing attention directed toward the region. In line with our analysis, we restricted our query to articles written in English that were published between 2002 and 2019 (For the full query, see Supplementary Materials, \FigScopusQuery{}). 

A search of the Scopus database, which draws from the same database as ICSR, using the query above, conducted in January 2025, yielded 3,458 articles. The same country identification algorithm was applied to extract the country names mentioned in the titles and abstracts of these articles, and 67.8\% of the articles mentioned at least one country.

\subsection*{Scholarly migration dataset}
To study the migration of researchers from and to countries affected by the Arab Spring, we used the Scholarly Migration Database \cite{akbaritabar2024bilateral, ScholarlyMigrationDatabase2022}. This dataset identifies changes in researchers' affiliated countries based on their publications and interprets them as scholarly migration, providing country-year-level statistics to analyze global talent flow. It utilizes bibliographic data from Scopus to identify migration events at the researcher level. To ensure accurate identification, the dataset relies on author disambiguation provided by Scopus, which reports a precision rate of 98.1\% at the researcher level.

We use the net migration rate (\( NMR_{i,t} \)) to quantify the relative balance of scholar migration between countries. It normalizes net migration by the total scholar population and is defined as
\[
NMR_{i,t} = \frac{I_{i,t} - E_{i,t}}{N_{i,t}},
\]

where \( I_{i,t} \) and \( E_{i,t} \) represent the number of scholars entering and leaving country \( i \) in year \( t \), respectively, and \( N_{i,t} \) is the total scholar population. This metric reflects the relative migration a country experiences in proportion to the size of its academic workforce, where a positive value indicates net brain gain (more researchers are entering than leaving), a negative value signifies brain drain (more researchers are leaving than entering), and the magnitude represents the relative degree of migration imbalance. 

\subsection*{Difference-in-Differences for scholarly attention panel data}

To identify the Arab Spring's impact on scholarly attention and funding, we adopt a Difference-in-Differences (DiD) statistical approach typically used to mimic an experimental research design when only observational data is available. This approach enables us to estimate the average treatment effect of the Arab Spring by comparing temporal changes in attention/funding to target countries (treatment group) with those observed in appropriate control groups, namely, the rest of the world, unaffected MENA countries, or matched countries identified through coarsened exact matching. For more detailed information on the matching procedure, refer to the Supplementary Text in the supplementary materials, \TabCEMCountries{} for the list of countries and \FigCEMBalance{} for checking the distribution of covariates for target and control groups.

The main DiD regression model is specified as:  
\[
\log y_{i}(t) = \sum_{s \in S} \beta_s \cdot \text{Post}_t \cdot \text{Treatment}^s_i +\lambda_t +\gamma_i  + X_{it}\theta + \varepsilon,
\]
where \( y_i(t) \) represents the outcome variable of interest, such as the logarithm of total, domestic, or foreign scholarly attention received by country \( i \) at time \( t \). The coefficient of interest, \( \beta_s \), captures the average treatment effect of the Arab Spring for treatment subgroup \( s \) (e.g., civil war (CW)), and $S$ is the set of all treatment subgroups. The variables \( \text{Post}_t \) and \( \text{Treatment}^s_i \) are binary indicators denoting the post-treatment period and whether country \( i \) belongs to the treatment group \( s \), respectively. Country fixed effects (\( \gamma_i \)) account for time-invariant heterogeneity across countries, and year fixed effects (\( \lambda_t \)) control for global temporal shocks. \( X_{it} \) represents a vector of time-varying country-level covariates (logarithms of GDP per capita, number of researchers, and total population). Standard errors are clustered at the country level to account for serial correlation. All the reported regressions exhibit statistically significant F-statistics, with \( p\text{-values} < 0.05 \). 

We also report the within \( R^2 \), a goodness measure of fit, which reflects the proportion of variation in the outcome explained by the model within countries over time, excluding between-country variation, and is calculated as \cite{verbeek2017guide}:
\[
R^2_{\text{within}} = \text{Corr}^2[\hat{y}_{it} - \hat{y}_i, \, y_{it} - \bar{y}_i],
\]
where \( y_{it} \) and \( \hat{y}_{it} \) are the observed and predicted outcomes for country \( i \) at time \( t \), and \( \bar{y}_i \) and \( \hat{y}_i \) are their respective country-level averages over time. 
To verify the validity of the regression results, we also conducted a parallel prior trends test to verify the parallel trends assumption \cite{huntington2018chapter}, and an examination of multicollinearity by ensuring that the Variance Inflation Factor (VIF) for the post-treatment variable remained below the conventional threshold of 5 \cite{sheather2009modern}, which was satisfied for all regressions.

The assumption of parallel trends is fundamental to difference-in-differences (DiD) designs. It states that, in the absence of treatment, treated and control countries would have followed similar trajectories over time. While this assumption is inherently untestable since the counterfactual is unobservable, it is possible to look for deviations from it before the treatment to provide support for it over the entire period \cite{huntington2018chapter}. We test whether the interaction term between time and treatment is significant before treatment, specifically \(2002\leq t <2011\), using the following regression model
\[
\log y_{i}(t) = \sum_{s \in S}\beta^p_s \cdot (t - 2011) \cdot \text{Treatment}^s_i + \lambda \cdot (t - 2011) + X_{it}\theta + \gamma_i + \varepsilon,
\]
where the parameter \(\beta^p_s\) captures the difference in trends between treated and control groups before the treatment period. The parameter \(\lambda\) reflects the overall time trend common to all countries, while \(\theta\) measures the effect of the time-varying covariates \(X_{it}\). The term \(\gamma_i\) accounts for country-specific fixed effects, capturing characteristics that are constant over time but vary across countries. 
The test's null hypothesis (\(\beta^p_s = 0\)) indicates no difference in pre-treatment trends between treated and control countries for subgroup \(s \in S\). The alternative hypothesis (\(\beta^p_s \neq 0\)) implies a violation of the parallel trends assumption. We conduct the tests at a significance level of \(\alpha = 0.05\). When the null hypothesis is rejected ($p<\alpha$), it suggests that treated and control countries exhibited different trends before treatment, which undermines the validity of the DiD framework. When the alternate hypothesis is rejected ($p\geq\alpha$), it does not prove that the assumption of parallel trends holds, but it provides statistical support for it. We find that the null hypothesis is only rejected for one coefficient in Table \ref{tab:attention_DiD} (indicated with a dagger) and in 8 coefficients in Table S5 in Supplementary Materials. We report all $p$-values for these tests in \TabPTAttention{} and \TabPTFunding{} in the Supplementary Materials.

\section*{Acknowledgments}

 We deeply thank the International Center for the Study of Research (ICSR) at Elsevier for providing us access to the Scopus dataset and its computing resources. We sincerely thank Emilio Zaghani and Ali Akbaritabar for their invaluable assistance in providing the scholarly migration database and sharing insights about its framework. We are deeply thankful to Daniel Acuna and Stephen V. David for creating the opportunity through the Science of Science Summer School 2022, where the conceptual seeds of this work were planted, and our collaborative research group took shape.  We also thank Dorian Quelle, Samuel Koovely, Yuan Zhang, and Reinhard Furrer at the University of Zurich for their valuable feedback. Finally, we acknowledge Juven Villacastin and Mimi Byun for their support during the development of the Scholarly Attention Project. 
 
\noindent \textbf{Funding:} Yasaman Asgari thanks the University of Zurich and the Digital Society Initiative for partially funding this project.

 \noindent\textbf{Author contributions}: Conceptualization: Y.A. and A.B. . Data curation: Y.A. and H.Z. Formal analysis: Y.A., H.Z., A.B. Methodology: Y.A., H.Z., A.B., R.R., O.O., M.S. Writing: Y.A., A.B., H.Z., R.R., O.O., M.S.
 
\noindent\textbf{Competing interests}: The authors declare they have no competing interests.
 
    \noindent \textbf{Data and materials availability}: All data and codes needed to reproduce our results have been deposited on the Harvard Dataverse at \url{https://doi.org/10.7910/DVN/9HXJJV}.




\clearpage

\section*{Supplementary Materials for \textit{The Arab Spring’s Impact on Science through the Lens of Scholarly Attention, Funding, and Migration}}

\begin{center}
Yasaman Asgari$^{\ast}$, Hongyu Zhou, Özgür Kadir Özer, Rezvaneh Rezapour, Mary Ellen Sloane, and Alexandre Bovet$^{\ast}$
$^\ast$ \{Yasaman.asgari, Alexandre.bovet\}@uzh.ch

\end{center}

\renewcommand{\thefigure}{S\arabic{figure}}
\renewcommand{\thetable}{S\arabic{table}}

\setcounter{figure}{0}
\setcounter{table}{0}

\section*{Supplementary Text}

\subsection*{Analysis of hyperprolific authors}

Authorship plays a crucial role in scientific research, and contributing to a paper should reflect meaningful participation in conducting the study. However, a growing concern has emerged regarding researchers who publish at exceptionally high rates.  Some scholars, referred to as hyperprolific researchers, author more than 72 papers per year \cite{ioannidis2018thousands} (equivalent to one paper every five days), while highly prolific researchers publish more than 60 papers annually \cite{conroy2024surge} (one paper every six days).

Recent investigations of this phenomenon have studied so-called hyperprolific researchers" who author more than 72 papers per year \cite{ioannidis2018thousands} (equivalent to one paper every five days), or "extremely productive researchers" publishing more than 60 papers annually \cite{conroy2024surge} (one paper every six days). They have shown that hyperprolific researchers are predominantly active in physics,  clinical medicine, technology, and engineering. We have excluded physics and engineering subject areas for separate reasons. The majority of hyperprolific authors are based in the United States, followed by Germany and Japan \cite{ioannidis2018thousands}.  Malaysia and Saudi Arabia—countries where publication incentives, including cash rewards, are also among the top countries with hyperprolific researchers \cite{abritis2017cash}. Notably, the rise of extremely productive researchers in Saudi Arabia has raised concerns about the potential use of questionable methods to increase publication counts \cite{conroy2024surge}.

Given the possible influence of hyperprolific researchers on scholarly attention and that Malaysia, Saudi Arabia, the USA, and Germany rank among the highest contributors in our dataset, we analyzed to ensure that their presence does not unduly affect our findings. While we have no direct evidence of misconduct by these researchers, we aim to assess their impact on our work.

By identifying authors of papers that mention at least one target country in their abstract or title, we found a total of 135,227 authors. We then analyzed their publication records during the study period and found 18 hyperprolific researchers—defined as those publishing more than 72 papers per year. Among these hyperprolific authors, three are from France, two each from Germany and the UK, one from Saudi Arabia, and one from the USA.
In comparison, within our entire dataset of 25 million publications, there are 69 hyperprolific researchers among 14,084,678 authors across 22 countries. The United States has the highest number, with 16 hyperprolific researchers, followed by the Netherlands (7), the UK (6), China, Finland, and Germany, each with 4.

Authors mentioning a target country account for less than 1\% of researchers globally, while they represent 26.1\% of all high-profile researchers. However, given the absence of Malaysia and the presence of only one high-profile researcher from the USA and Saudi Arabia, we conclude that the impact of high-profile researchers in our study is minimal.

\subsection*{Topic analysis of Arab Spring–related research}
To examine the evolution of Arab Spring-related research topics within each country group, we selected articles published after 2011 that mentioned at least one country from the respective group. The resulting corpora include 856, 560, and 356 documents for the GO, CW, and GC groups, respectively, which provide sufficient data for topic modeling. Notably, 201 documents mentioned both GO and CW countries, 148 mentioned both GO and GC countries, and 107 mentioned both CW and GC countries. We applied BERTopic using a count vectorizer (n-grams of length 1–3) with English stop-words. Topics were modeled separately within each group to capture group-specific dynamics, and an additional model was fitted across all documents to identify shared themes. The topics Turkey’s Regional Role, Gulf Monarchies, Refugees, Social Media, Islamist Parties, Women’s Rights, and Oil \& Financial Markets, EU policy \& migration appear consistently across both approaches, refer to tables \ref{three_models_topics} and \ref{one_model_topics}. Therefore, in our analysis, we focus on the evolution of these shared topics after the Arab Spring within the three country groups (GO, CW, and GC).

\subsection*{Coarsened Exact Matching for Difference in Difference}
To assess the causal impact of the Arab Spring on scholarly attention and research funding, we employed a Difference-in-Differences (DiD) design. A central challenge in such designs is ensuring that treatment (in this case, the target countries) and control groups are sufficiently comparable in the pre-treatment period.

Therefore, we employed coarsened exact matching (CEM) \cite{iacus2012causal} to identify a more suitable control group. We utilized the Python implementation of the CEM\cite{iacus2009cem} at the country level to construct a balanced control group. First, we restricted the data to the pre-event period (years $<2011$) and computed the mean values of scholarly attention (log-transformed), GDP per capita (log), population (log), and researcher population (log) when needed. Each continuous covariate was standardized into a z-score and then coarsened into five bins using fixed cutpoints $(-\infty,\,-3,\,-1.5,\,-0.75,\,0.75,\,1.5,\,3,\,\infty)$. Strata were defined as the joint combination of the bins across all covariates. We retained only those strata containing at least one treated and one control country. The resulting matched set is represented in Table \ref{coarsened_countries}.
\section*{Supplementary Tables}
\begin{table}[h!]
\centering
\begin{tabular}{|c|r|r|}
\hline
\textbf{Country Code} & \textbf{2002} & \textbf{2019} \\
\hline
Bahrain & 710,553 & 1,483,756 \\
Kuwait & 2,059,974 & 4,442,316 \\
Oman & 2,337,140 & 4,591,241 \\
Libya & 5,508,410 & 6,951,033 \\
Jordan & 5,616,066 & 10,671,891 \\
Tunisia & 9,971,515 & 11,875,081 \\
Syria & 17,468,332 & 20,353,534 \\
Yemen & 20,835,086 & 35,111,408 \\
Morroco & 29,198,142 & 36,210,898 \\
Egypt & 76,239,137 & 107,553,158 \\
World's Median & 8,400,180&9,054,000\\
\hline
\end{tabular}
\caption{Population data from 2002 and 2019 for selected countries}
\label{SI-tab:population_data}
\end{table}
\newpage

\begin{table}[!ht]
    \centering
    \begin{tabular}{ll}
        \toprule
        \textbf{Major Discipline} & \textbf{Subject Areas} \\
        \midrule
        \multirow{5}{*}{Life Sciences} & Agricultural and Biological Sciences \\
        & Biochemistry, Genetics and Molecular Biology \\
        & Immunology and Microbiology \\
        & Neuroscience \\
        & Pharmacology, Toxicology, and Pharmaceutics 	\\
        \midrule
        \multirow{10}{*}{Physical Sciences} & Chemical Engineering \\
        & Chemistry \\
        & Earth and Planetary Sciences \\
        & Energy \\
        & Environmental Science \\
        &Materials Science\\
        & \textcolor{red}{*Computer Science}\\
        & \textcolor{red}{*Engineering }\\
        & \textcolor{red}{*Mathematics }\\
        & \textcolor{red}{*Physics and Astronomy }\\
        \midrule
        \multirow{6}{*}{Health Sciences} & Medicine \\
        & Nursing \\
        & Veterinary \\
        & Dentistry \\
        & Health Professions \\
        \midrule
        \multirow{7}{*}{Social Sciences \& Humanities} & Arts and Humanities \\
        & Business, Management and Accounting \\
        & Decision Sciences \\
        & Economics, Econometrics and Finance \\
        & Psychology \\
        & Social Sciences \\
        \midrule
         \multirow{1}{*}{ \(-\)}& \textcolor{red}{*Multidisciplinary}\\
        \bottomrule
    \end{tabular}
    \caption{\textbf{Scopus subject areas and their major disciplines.} We excluded the subject areas displayed in red and articles classified as multidisciplinary.}
    \label{tab: scopus}
\end{table}

\begin{table}[ht]
\centering
\begin{adjustbox}{max width=\textwidth}
\begin{tabular}{|l|c|l|c|l|c|}
\hline
\textbf{Government overthrow} & \textbf{Count} & 
\textbf{Civil wars} & \textbf{Count} & 
\textbf{Governmental changes} & \textbf{Count} \\
\hline
Arab Spring              & 304 & Syria/Libya/Yemen              & 313 & Morocco                        & 127 \\
Egypt/Tunisia            & 253 & Arab Spring                     & 132 & Arab Spring                    & 103 \\
Social Media             & 146 & Social Media                    &  50 & Social Media                   &  48 \\
Oil \& Financial Markets &  50 & Turkey’s Regional Role          &  41 & Gulf Monarchies                &  29 \\
Women’s Rights           &  41 & Refugees                        &  14 & Palestinian Refugees in Jordan &  24 \\
EU Policy \& Migration   &  27 & EU Security Agenda \& Migration &  10 & Oil \& Financial Markets       &  14 \\
Turkey’s Regional Role   &  20 &                                 &     & Kuwait                         &  11 \\
Inequality, Employment   &  15 &                                 &     &                                &     \\
\hline
\end{tabular}
\end{adjustbox}
\caption{Topics and their frequencies across the three country groups (GO, CW, GC), based on one topic model per group.}
\label{three_models_topics}
\end{table}
\newpage

\begin{table}[ht]
\centering
\begin{adjustbox}{max width=0.9\textwidth}
\begin{tabular}{|l|c|c|c|}
\hline
\textbf{Topic} & \textbf{Government overthrow} & \textbf{Civil wars} & \textbf{Governmental changes} \\
\hline
Arab Spring                                     & 256 & 163 & 113 \\
Egypt                                           & 171 &  6  &  4  \\
Tunisia                                         &  38 &  7  &  7  \\
Libya                                           &  13 & 85  &  1  \\
Syria                                           &   3 & 66  &  4  \\
Yemen                                           &   1 & 21  &     \\
Morocco                                         &  13 &     & 40  \\
Jordan                                          &   3 &  6  & 21  \\
Kuwait                                          &   4 &  3  & 34  \\
Gulf monarchies: Saudi                          &   7 & 13  & 23  \\
Russia \& Iran                                  &   3 & 17  &  2  \\
Refugees                                        &   5 & 18  &  9  \\
Social Media                                    & 143 & 60  & 36  \\
Oil \& Financial Markets                        &  44 & 15  & 18  \\
Women’s Rights                                  &  43 &  3  &  9  \\
Islamist parties                                &  36 &  4  &  5  \\
Tourism                                         &  11 &  3  &  8  \\
Military                                        &  10 & 11  &  6  \\
Employment, Education, and Economy              &  10 &  1  &  2  \\
EU Policy \& Migration                          &  22 &  8  & 11  \\
Water Scarcity, Climate Change, and Resources   &   3 &  9  &  3  \\
Turkey’s Regional Role                          &  17 & 41  &    \\
\hline
\end{tabular}
\end{adjustbox}
\caption{Frequencies of topics across the three country groups using one unified topic model for all Arab Spring-related research.}
\label{one_model_topics}
\end{table}

\newpage

\begin{table}[ht]
    \centering
    \begin{adjustbox}{max width=\textwidth}
    \begin{tabular}{lcccccccc}
        \toprule
        \textbf{Model} & \textbf{SampleGroup} & \textbf{Accuracy Union} & \textbf{Accuracy Intersection} & \textbf{Jaccard Union} & \textbf{Jaccard Intersection} \\
        \midrule
        \multirow{3}{*}{Our Algorithm} & with mention arab & \textbf{0.840} & 0.750 & 0.905 & 0.821 \\
        & with mention & \textbf{0.929} &\textbf{ 0.775} & \textbf{0.946} & \textbf{0.804} \\
        & field 20 &\textbf{ 0.973} & \textbf{0.960} & \textbf{0.975} & \textbf{0.961} \\
        \midrule
        \multirow{3}{*}{ChatGPT 4o} & with mention arab & 0.840 & \textbf{0.790} &\textbf{0.914} & \textbf{0.855} \\
        & with mention & 0.803 & 0.728 & 0.845 & 0.778 \\
        & field 20 & 0.93 & 0.907 & 0.934 & 0.907\\
        \midrule
        \multirow{3}{*}{BERT-base-NER} & with mention arab & 0.605 & 0.570 & 0.786 & 0.735 \\
        & with mention & 0.632 & 0.582 & 0.715 & 0.669 \\
        & field 20 & 0.896 & 0.894 & 0.908 & 0.905 \\
        \midrule
        \multirow{3}{*}{BERT-large-NER} & with mention arab & 0.620 & 0.58 & 0.797 & 0.746 \\
        & with mention & 0.632 & 0.582 & 0.719 & 0.675 \\
        & field 20 & 0.908 & 0.910 & 0.921 & 0.920 \\
        \bottomrule
    \end{tabular}
    \end{adjustbox}
    \label{tab:performance_metrics}
        \caption{Performance Metrics Comparison}

\end{table} 

\newpage

\begin{table}[ht]
\centering
\begin{adjustbox}{max width=\textwidth}
\begin{tabular}{|p{2.5 cm}||p{2cm}|p{6cm}|p{7.5cm}|}
\hline
\textbf{Group} &\textbf{Attention Type} & \textbf{Matched Covariates} & \textbf{Matched Countries} \\
\hline
CEM group $^{(1)}$
&Total attention 
& Pre-Arab Spring average log total attention, average log GDP per capita, average log population, average log researcher population 
& \textbf{Angola}, \textbf{Bulgaria}, \textbf{Bolivia}, \textbf{Cameroon}, \textbf{Costa Rica}, \textbf{Cuba}, \textbf{Dominican Republic}, \textbf{Algeria}, \textbf{Ecuador}, \textbf{Honduras}, Haiti, \textbf{Iraq}, \textbf{Kazakhstan}, Kyrgyzstan, Laos, \textbf{Lebanon}, \textbf{Sri Lanka}, Malaysia, \textbf{Peru}, \textbf{Puerto Rico}, \textbf{Paraguay}, \textbf{Romania}, Rwanda, \textbf{Senegal}, El Salvador, \textbf{Slovenia}, Togo, \textbf{Trinidad and Tobago}, \textbf{Ukraine}, \textbf{Venezuela} \\
\hline
CEM group $^{(2)}$
&Domestic attention 
& Pre-Arab Spring  average log domestic attention, average log GDP per capita, average log researcher population 
& \textbf{Angola}, Albania, \textbf{Bulgaria}, Bosnia and Herzegovina, Belarus, \textbf{Bolivia}, Botswana, \textbf{Cameroon}, Colombia, \textbf{Costa Rica}, \textbf{Cuba}, Cyprus, \textbf{Dominican Republic}, \textbf{Algeria}, \textbf{Ecuador}, Estonia, Gabon, Guatemala, \textbf{Honduras}, Croatia, Indonesia, \textbf{Iraq}, Jamaica, \textbf{Kazakhstan}, Cambodia, \textbf{Lebanon}, \textbf{Sri Lanka}, Lithuania, Latvia, North Macedonia, Mozambique, Nicaragua, \textbf{Peru}, Philippines, \textbf{Puerto Rico}, \textbf{Paraguay}, Palestine, \textbf{Romania}, Saudi Arabia, \textbf{Senegal}, El Salvador, Slovakia, \textbf{Slovenia}, \textbf{Trinidad and Tobago}, \textbf{Ukraine}, Uruguay, \textbf{Venezuela} \\
\hline
CEM group $^{(3)}$
&Foreign attention 
& Pre-Arab Spring  average log foreign attention, average log GDP per capita, average log population 
& \textbf{Angola}, Azerbaijan, \textbf{Bulgaria}, Belarus, \textbf{Bolivia}, \textbf{Cameroon}, Colombia, \textbf{Costa Rica}, \textbf{Cuba}, \textbf{Dominican Republic}, \textbf{Algeria}, \textbf{Ecuador}, Guatemala, \textbf{Honduras}, Haiti, Iran, \textbf{Iraq}, \textbf{Kazakhstan}, Kyrgyzstan, Laos, \textbf{Lebanon}, \textbf{Sri Lanka}, Malaysia, \textbf{Peru}, Nicaragua, Philippines, Poland, \textbf{Puerto Rico}, \textbf{Paraguay}, \textbf{Romania}, \textbf{Senegal}, Sierra Leone, El Salvador, \textbf{Slovenia}, Chad, Tajikistan, \textbf{Trinidad and Tobago}, \textbf{Ukraine}, Uzbekistan, \textbf{Venezuela} \\
\hline
\end{tabular}
\end{adjustbox}
\caption{Coarsened Exact Matching (CEM) country matches across attention types. Shared countries across all three groups are shown in bold.}
\label{coarsened_countries}
\end{table}

\newpage

\begin{table}[!ht]
\centering
\resizebox{\textwidth}{!}{%
\begin{tabular}{l|c|c|
>{\columncolor{green!15}}c|
>{\columncolor{blue!15}}c
>{\columncolor{blue!15}}c
>{\columncolor{blue!15}}c}
\toprule

\textbf{Dependent Variable} & \textbf{Control Variables} & \textbf{Control group} & \textbf{\(p_{\text{Target}}\)} & \textbf{\(p_{\text{GO}}\)} & \textbf{\(p_{\text{CW}}\)} & \textbf{\(p_{\text{GC}}\)} \\
\midrule
Log Attention               &\begin{tabular}{@{}c@{}} log GDP per capita, log pop.   \\log Rpop. \end{tabular}& Rest of world&0.635& 0.175 & 0.855&0.612\\
&  & Rest of MENA &0.990&0.284 &  0.599&0.600\\
&  & CEM group$^{(1)}$ &0.092&0.014 &  0.317 &0.350\\
\midrule
Log Domestic Attention               & log GDP per capita, log Rpop. & Rest of world& 0.915&  0.042 & 0.096&0.202\\
&  & Rest of MENA &0.661& 0.087 & 0.452&0.891\\
&  & CEM group$^{(2)}$ &0.104&0.816&  0.001&0.0736\\

\midrule

Log Foreign Attention               & log GDP per capita, log pop.   & Rest of world&0.933 & 0.581& 0.243&0.436\\
& & Rest of MENA &0.255& 0.441 & 0.684&0.181\\

&  & CEM group$^{(3)}$ &0.2383&0.001&  0.007&0.465\\
\bottomrule
\end{tabular}%
}
\caption{\textbf{\(p-\)values for pre-treatment parallel trend for scholarly attention} Models control for log GDP, population, and research population where applicable. Each colored column represents $p$-values from separate regressions comparing the target countries with two control groups: the rest of the world and MENA countries less affected by the Arab Spring.}
\label{SI-tab:DiD_results_p_values_attention}

\end{table}

\begin{table}[!ht]
\centering
\resizebox{\textwidth}{!}{%
\begin{tabular}{l|c|c|
>{\columncolor{green!15}}c|
>{\columncolor{blue!15}}c
>{\columncolor{blue!15}}c
>{\columncolor{blue!15}}c}
\toprule

\textbf{Dependent Variable} & \textbf{Control Variables} & \textbf{Control group} & \textbf{\(p_{\text{Target}}\)} & \textbf{\(p_{\text{GO}}\)} & \textbf{\(p_{\text{CW}}\)} & \textbf{\(p_{\text{GC}}\)} \\
\midrule
Log Funding               &\begin{tabular}{@{}c@{}} log GDP per capita, log pop.\end{tabular}& Rest of world&0.124& 0.358 & 0.447&0.002\\
&  & Rest of MENA &0.019& 0.748  & 0.091&0.001 \\
\midrule
Log Domestic Funding               & log GDP per capita& Rest of world &0.999&  0.158 & 0.000&0.923\\
&  & Rest of MENA &0.586& 0.343& 0.005& 0.629\\
\midrule

Log Foreign Funding               & log GDP per capita, log pop.   & Rest of world&0.125&  0.263 &0.670&0.005 \\
& & Rest of MENA& 0.017&0.439   & 0.182&0.001\\
\bottomrule
\end{tabular}%
}
\caption{\textbf{\(p-\)values for pre-treatment parallel trend test for funding} Models control for log GDP, population where applicable. Each colored column represents $p$-values from separate regressions comparing the target countries with three control groups: the rest of the world and MENA countries less affected by the Arab Spring, and the countries matched through the coarsened exact matching }
\label{SI-tab:DiD_results_p_values_funding}

\end{table}

\clearpage

\begin{table}[!ht]
\centering
\resizebox{\textwidth}{!}{%
\begin{tabular}{l|c|c||
>{\columncolor{green!15}}c|
>{\columncolor{green!15}}c||
>{\columncolor{blue!15}}c
>{\columncolor{blue!15}}c
>{\columncolor{blue!15}}c|
>{\columncolor{blue!15}}c}
\toprule

\textbf{Dependent Variable} & \textbf{Control Variables} & \textbf{Control group} & \textbf{\(\beta_{\text{Target}}\)}& \(R^2 {\text{within}}\) & \textbf{\(\beta_{\text{GO}}\)} & \textbf{\(\beta_{\text{CW}}\)} & \textbf{\(\beta_{\text{GC}}\)} & \(R^2 {\text{within}}\)\\

\midrule
Log Funding               &\begin{tabular}{@{}c@{}} log GDP per capita, log pop.\end{tabular}& Rest of world& \begin{tabular}{@{}c@{}}-0.0121 \\ ( 0.119)\end{tabular}&0.1743& \begin{tabular}{@{}c@{}}0.2280 \\ (0.145)\end{tabular} & \begin{tabular}{@{}c@{}}0.3452$^{***}$ \\ (0.121)\end{tabular} &\begin{tabular}{@{}c@{}}-0.3248 $^{***,\dagger}$\\ (0.109)\end{tabular}&0.2064\\
&  & Rest of MENA & \begin{tabular}{@{}c@{}}-0.4361 $^{**,\dagger}$\\ (0.202)\end{tabular} &-0.1239& \begin{tabular}{@{}c@{}} -0.1251 \\ (0.222)\end{tabular} & \begin{tabular}{@{}c@{}} -0.1252$^{*}$ \\ (0.214)\end{tabular} &\begin{tabular}{@{}c@{}}-0.6587$^{***,\dagger}$  \\ (0.195)\end{tabular}&0.0146\\
\midrule
Log Domestic Funding               & log GDP per capita& Rest of world & \begin{tabular}{@{}c@{}} 0.1547 \\ (0.217)\end{tabular}&0.0511 & \begin{tabular}{@{}c@{}} 1.0938$^{***}$ \\ (0.112)\end{tabular} & \begin{tabular}{@{}c@{}} -0.4725$^{*,\dagger}$ \\ (0.245)\end{tabular} &\begin{tabular}{@{}c@{}}0.1081  \\ (0.128)\end{tabular}&0.0312\\
&  & Rest of MENA & \begin{tabular}{@{}c@{}} -0.4754 \\ (0.377)\end{tabular}& -0.1808 & \begin{tabular}{@{}c@{}}  0.4840 \\ (0.341)\end{tabular} & \begin{tabular}{@{}c@{}} -1.4139$^{***,\dagger}$ \\ ( 0.275)\end{tabular} &\begin{tabular}{@{}c@{}}-0.5466$^{*}$  \\ (0.313)\end{tabular}&-0.2873\\
\midrule

Log Foreign Funding               & log GDP per capita, log pop.   & Rest of world&\begin{tabular}{@{}c@{}}0.0299 \\ (0.103)\end{tabular}&0.1517 & \begin{tabular}{@{}c@{}}0.1518  \\ (0.169)\end{tabular} & \begin{tabular}{@{}c@{}}0.3658$^{***}$ \\ (0.104)\end{tabular} &\begin{tabular}{@{}c@{}} -0.2186 $^{**, \dagger}$ \\ ( 0.095 )\end{tabular}&0.1818\\
& & Rest of MENA &\begin{tabular}{@{}c@{}} -0.1209 $^{ \dagger}$ \\ ( 0.109 )\end{tabular}&-0.1841 & \begin{tabular}{@{}c@{}}-0.1298 \\ (0.188)\end{tabular} & \begin{tabular}{@{}c@{}} -0.0283 \\ (0.157)\end{tabular} &\begin{tabular}{@{}c@{}} -0.4046 $^{***, \dagger}$ \\ (0.127)\end{tabular}&-0.0777\\
\bottomrule
\end{tabular}%
}
\caption{\textbf{Difference in differences estimates of the impact of Arab Spring on funding.} Estimated coefficients from difference-in-differences regressions evaluating the post-treatment effects of Arab Spring on log funding (total, domestic, foreign). Models control for log GDP and population, where applicable. Each colored column represents estimates from separate regressions and compares the target countries with two control groups: the rest of the world and MENA countries less affected by the Arab Spring. Standard errors are in parentheses. ***, **, and * denote statistical significance at the 1\%, 5\%, and 10\% levels, respectively. $\dagger$ denotes cases where the pre-trend parallel assumption is violated with \(p-\)value of 5\%}
\label{SI-tab:DiD_funding}

\end{table}

\newpage
\section*{Supplementary Figures}
\begin{figure}[!h]
    \centering
    \includegraphics[width=\linewidth]{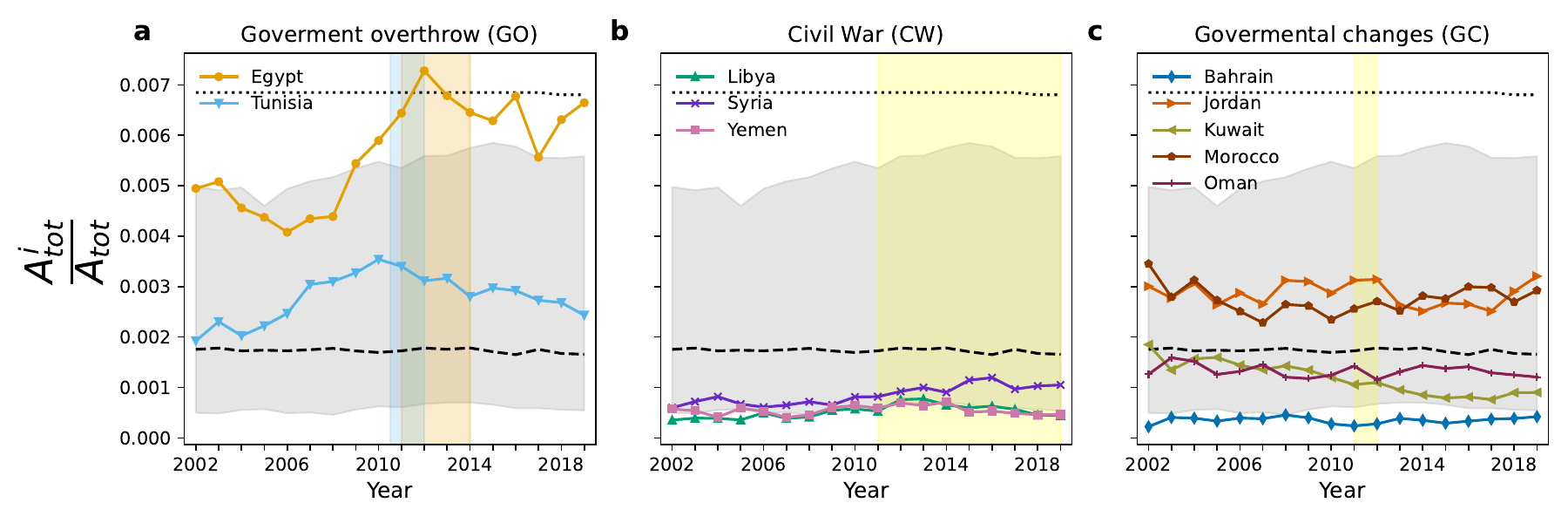}
    \caption{\textbf{Shifts in shares of scholarly attention to target countries.} Share of scholarly attention received by target countries, calculated as the attention to country \(i\) divided by the total global attention. In all panels, the gray area represents the interquartile range for global trends, with the dashed line indicating the
    median and the dotted line showing the mean.  Country-specific trends reveal significant increases in the share and distribution percentiles of scientific publications for Egypt (a) and Syria (b). The share peaks in 2012 for Egypt, 2010 for Tunisia, and 2016 for Syria. The governmental changes group trends (c) show decreases after the Arab Spring for Jordan and Kuwait.}
    \label{SI-share_attention}
\end{figure}

\begin{figure}[!ht]
    \centering
    \includegraphics[width=\textwidth]{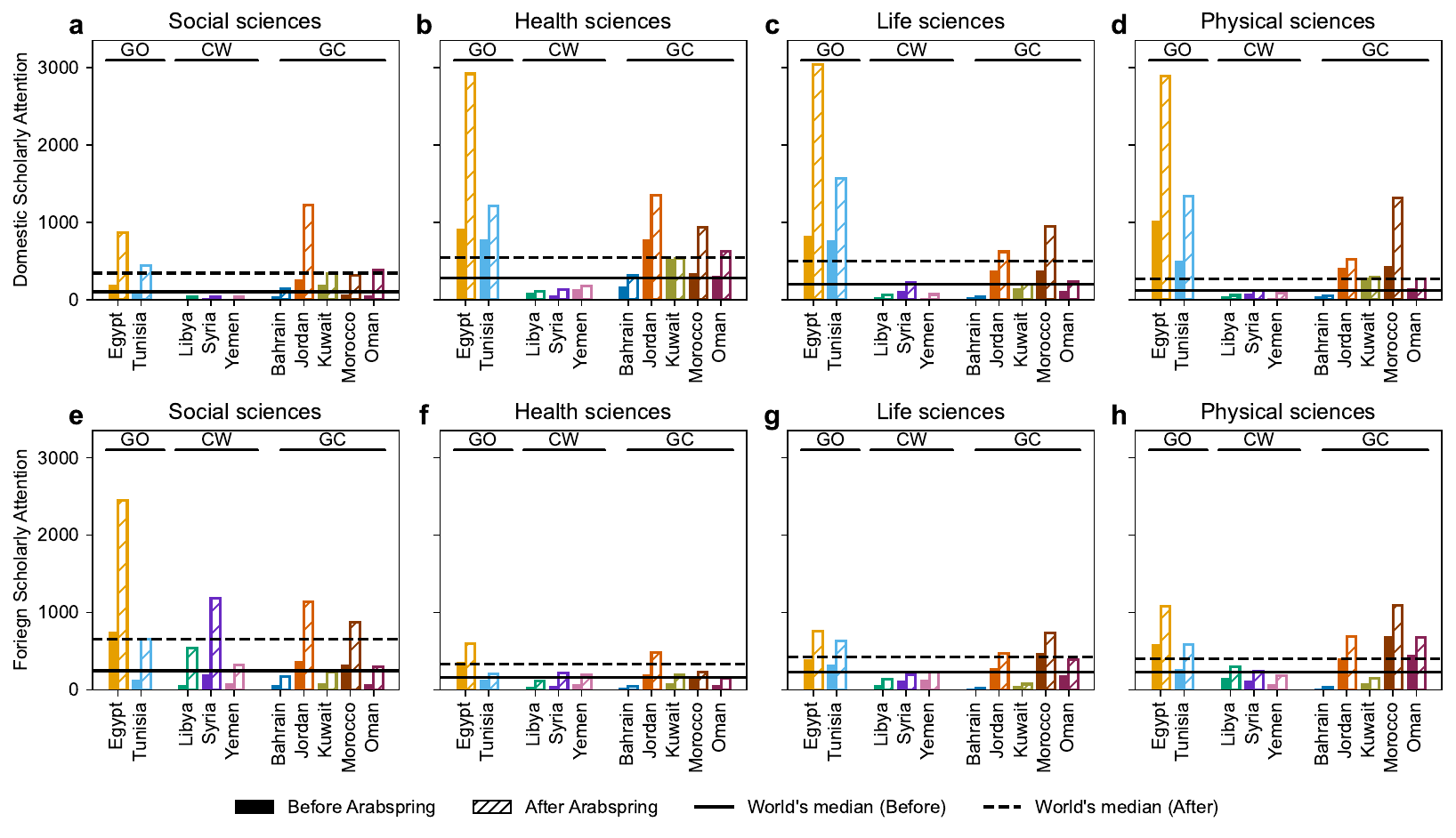}
    \caption{\textbf{Domestic and foreign scholarly attention before and after the Arab Spring by discipline.} The target countries are split into three groups: government overthrow (GO), civil war (CW), and governmental changes (GC). Filled bars represent values before the Arab Spring, while hatched bars indicate values after the Arab Spring. The world's median before and after the Arab Spring is marked with solid and dashed black lines, respectively. In domestic attention (\textbf{a–d}), social sciences consistently held the smallest values for most countries before the Arab Spring. After the Arab Spring, the social sciences' scholarly attention on governmental changes has increased. In foreign attention (\textbf{e–h}), social sciences showed notably higher values, particularly in Egypt, Jordan, Syria, and Morocco.}
    \label{SI-domestic_foreign_count}
\end{figure}

\begin{figure}[!ht]
    \centering
    \includegraphics[width=\textwidth]{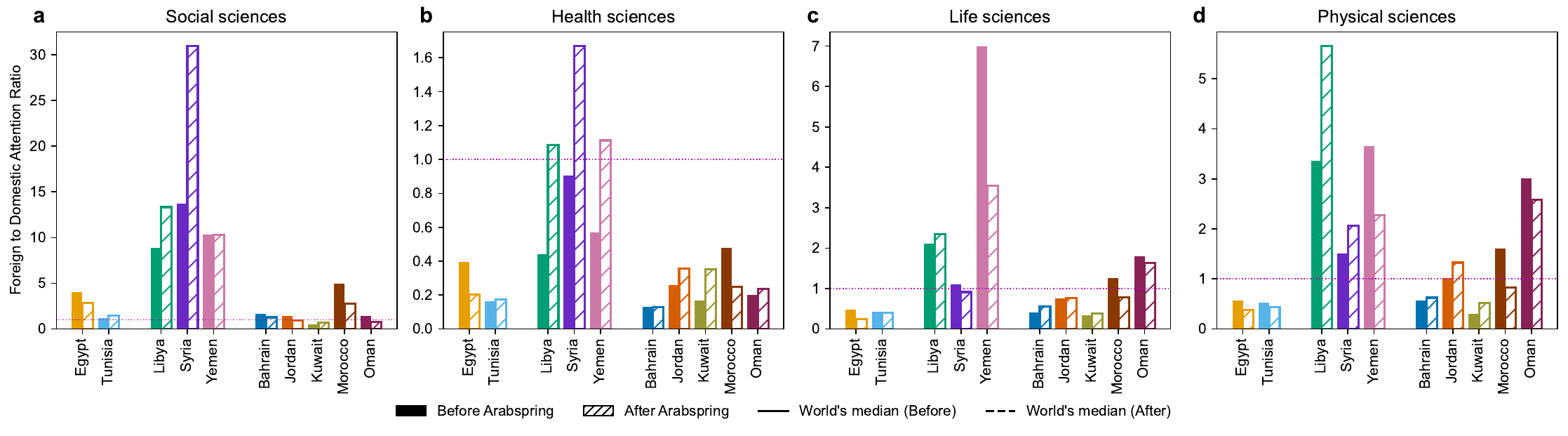}
    \caption{\textbf{Foreign to Domestic attention ratio before and after the Arab Spring by discipline.} Foreign-to-domestic attention ratios are calculated for the four main disciplines, (a) Social Sciences, (b) Health Sciences, (c) Life Sciences, and (d) Physical Sciences, before and after the Arab Spring, for each target country. In Egypt, domestic attention consistently surpassed foreign attention across all disciplines. In countries affected by civil war, the foreign-to-domestic attention ratio increased substantially in the Social Sciences, reaching very high values, and rose from below one to above one in Health Sciences. Meanwhile, the ratio for the governmental changes group displayed varied patterns across disciplines.}
    \label{SI-discipline_ratio}
\end{figure}

\begin{figure}
    \centering
    \includegraphics[width=0.7\linewidth]{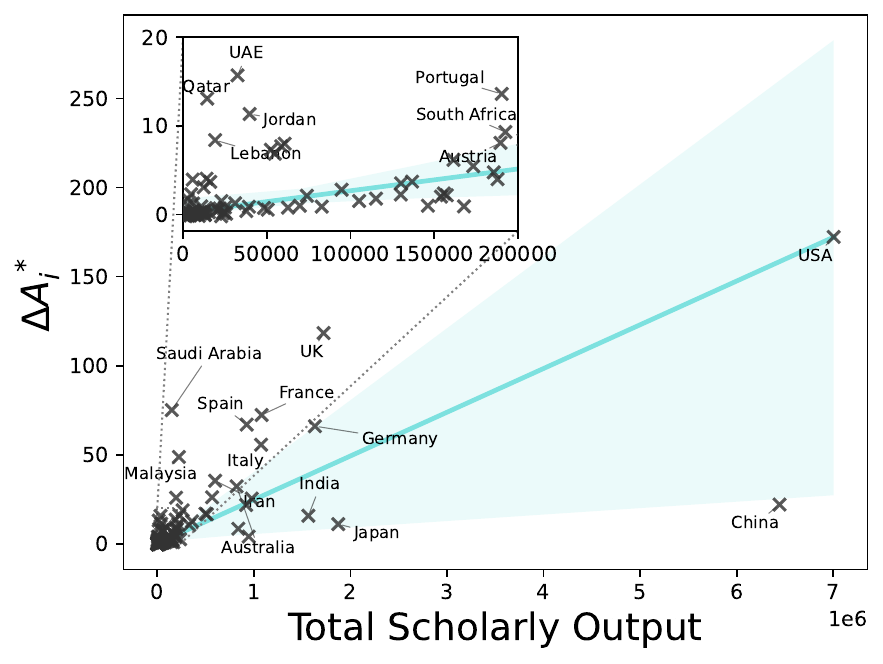}
    \caption{\textbf{Changes in scholarly attention to target countries relative to total scholarly output.} The increase in scholarly attention of sources countries strongly correlates with their total scholarly output (a proxy for research capacity and productivity) during the period (Pearson Correlation: 0.684; Spearman Correlation 0.845 both with \(p\text{-value} < 0.0001\)) The USA, Japan, and China have the highest outputs, but the USA shows a significantly greater surge in attention to target countries (172 more papers-fractionally counted-each year). The difference in target scholarly attention is 22 for China and 11 for Japan.}
    \label{SI-attention_vs_scholarlyoutput}
\end{figure}

\begin{figure}
    \centering
    \includegraphics[width=0.7\linewidth]{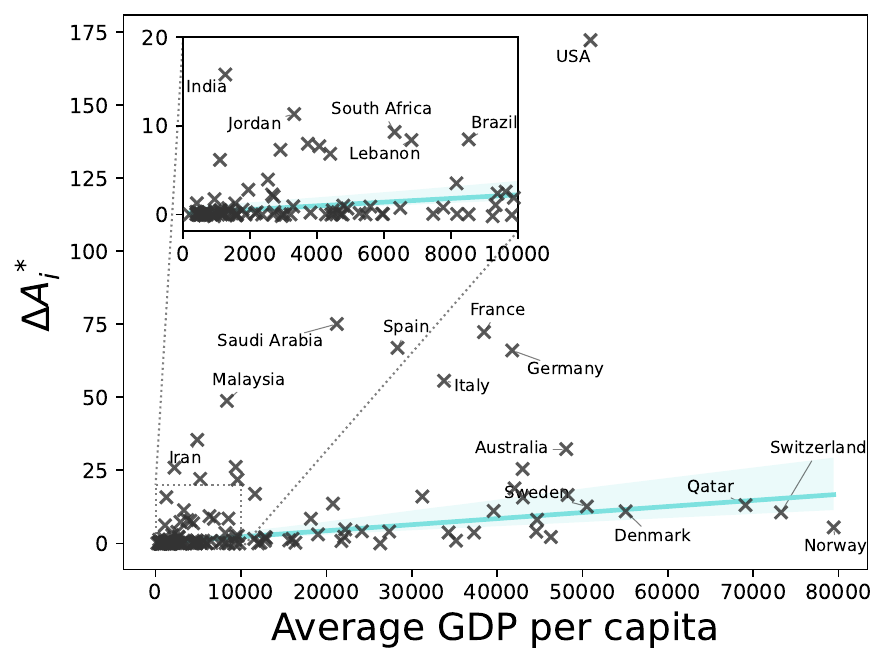}
    \caption{\textbf{Shifts in scholarly attention towards target countries as a function of the average GDP per capita.} The increase in the contribution of source countries to scholarly attention of target countries correlates with their average GDP per capita between 2002 and 2019 (Pearson Correlation: 0.432; Spearman Correlation 0.608, both with \(p\text{-value} < 0.0001\), suggesting that economic factors might explain the increased scholarly attention. The contributions of the USA, France, Germany, Italy, Saudi Arabia, and Malaysia to this attention exceeded the global average pattern.}
    \label{SI-attention_vs_gdp}
\end{figure}

\begin{figure}
    \centering
    \includegraphics[width=0.7\linewidth]{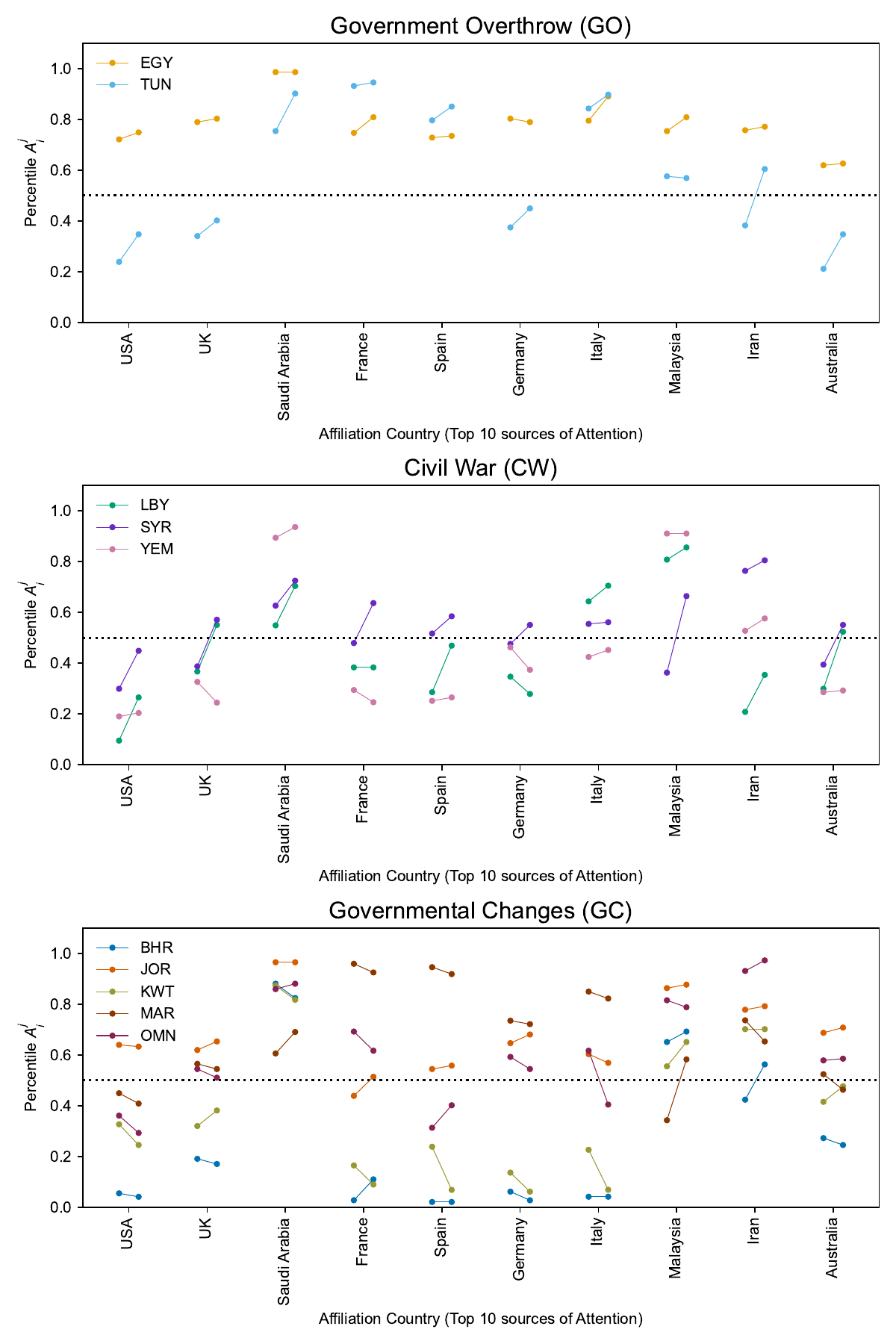}
\caption{\textbf{Rank changes in Scholarly attention to target Countries before and after the Arab Spring from the perspective of the top 10 sources of attention} Changes in the rank of scholarly attention from major sources to each target country, categorized into three subgroups: (a) Government Overthrown, (b) Civil War, and (c) Governmental Changes. The ranks for Syria, Libya, Tunisia, and Jordan generally increased after the Arab Spring. In contrast, Egypt’s rank gained slightly compared to the pre-Arab Spring period.}

\label{SI-rankchanges}

\end{figure}
\clearpage

\begin{figure}
    \centering
    \includegraphics[width=0.7\linewidth]{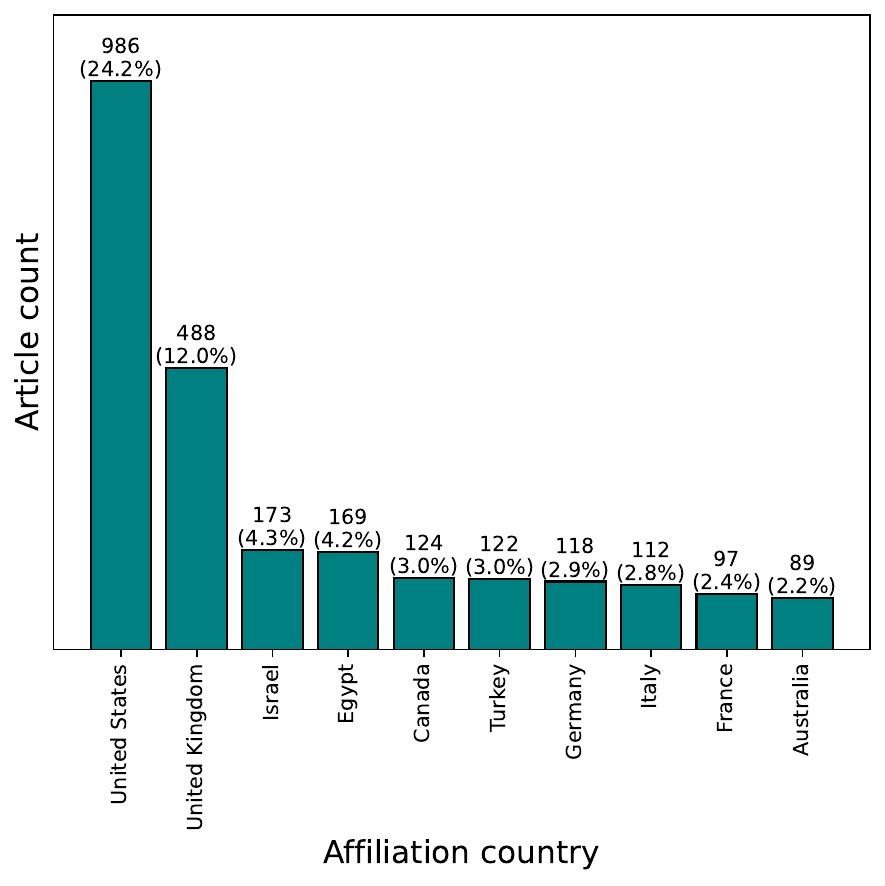}
    \caption{\textbf{Top contributing countries in Arab Spring research between 2002-2019.} The ten most frequent affiliation countries in Arab Spring-related research, along with their share and number of articles. The majority of contributions come from Western countries, including the USA, UK, France, Germany, Canada, Italy, and Spain. Turkey, Israel, and Egypt—countries directly or indirectly affected by the Arab Spring—are also among the top contributors.}
    \label{SI-AS_related_aff}
\end{figure}

\begin{figure}
    \centering
    \includegraphics[width=0.7\linewidth]{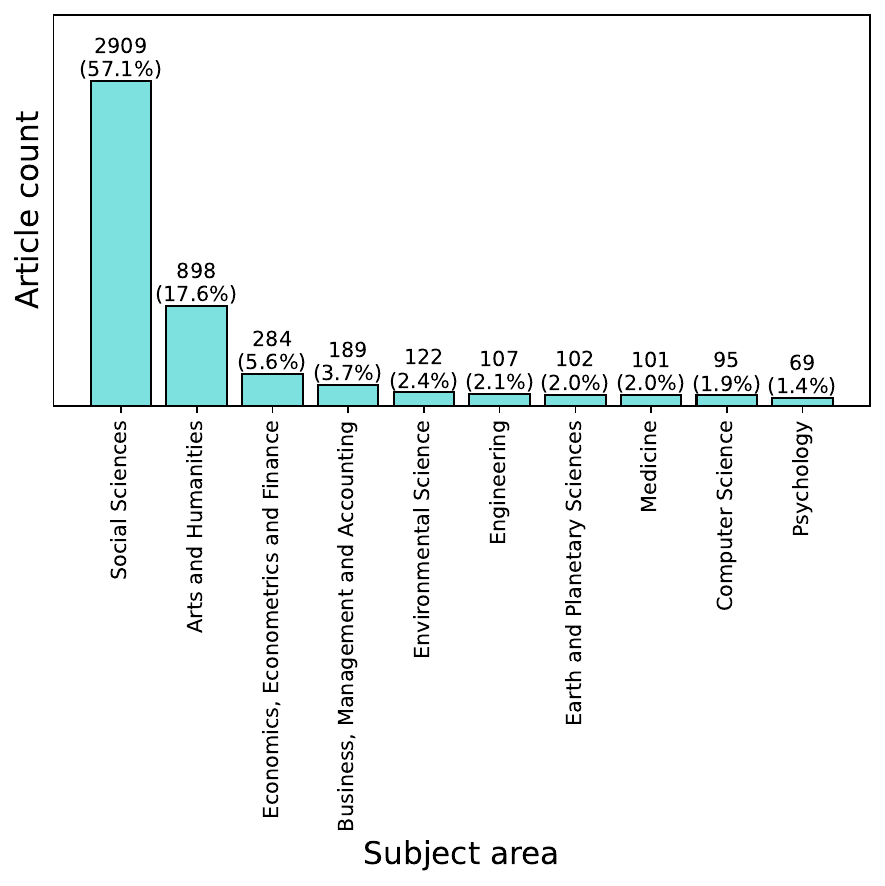}
\caption{\textbf{
Subject area Distribution in Arab Spring Research Between 2002–2019.} The top ten most frequent subject areas (out of 27) in Arab Spring-related research, presenting their respective share and number of articles. Most of these studies are concentrated in the social sciences, arts, and humanities.}
    \label{SI-AS_related_SA}
\end{figure}

\begin{figure}
    \centering
    \includegraphics[width=\linewidth]{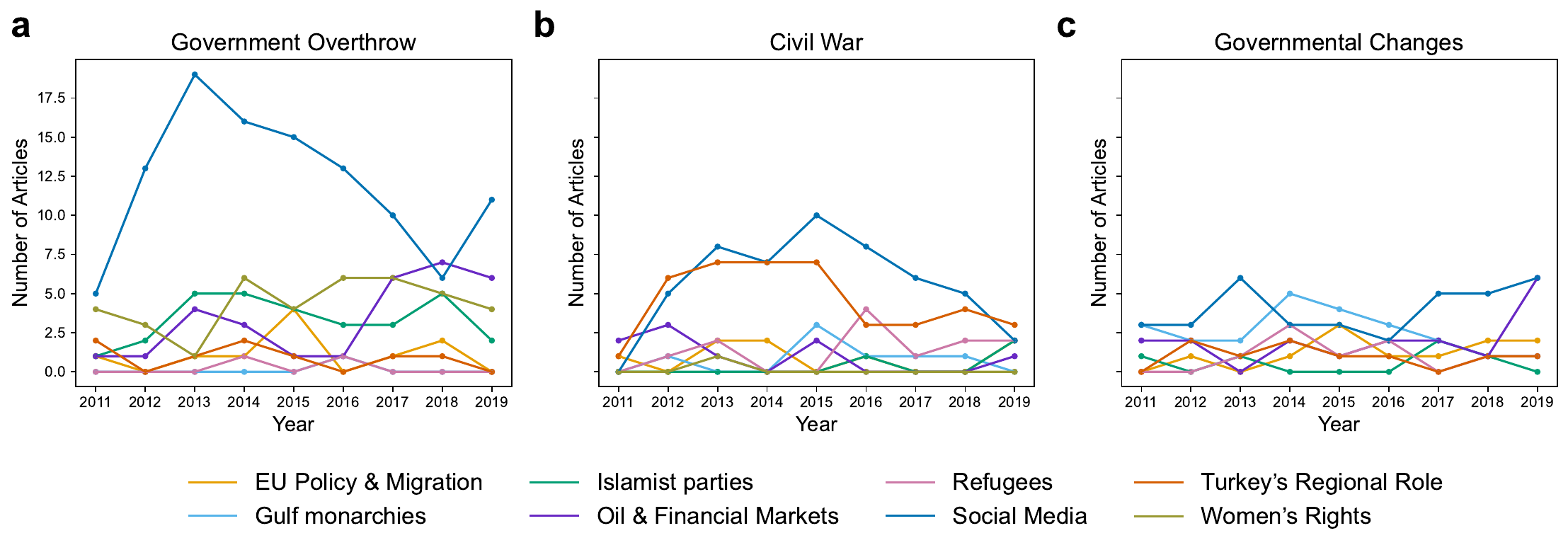}
\caption{\textbf{Topic dynamics across Arab Spring country groups.} Panels (a–c) show the annual topic counts and panels (d–f) the topic shares for scholarly attention covering different thematic areas between 2011 and 2019. Countries are grouped according to their outcomes following the Arab Spring: government overthrow (a, d), civil war (b, e), and governmental changes (c, f). Topic counts indicate the raw number of publications per year mentioning each theme. Prominent topics include social media, oil and financial markets, EU policy and migration, Islamist parties, Turkey’s regional role, and women’s rights.}
    \label{SI-Topic}
\end{figure}

\begin{figure}
    \centering
    \includegraphics[width=0.7\linewidth]{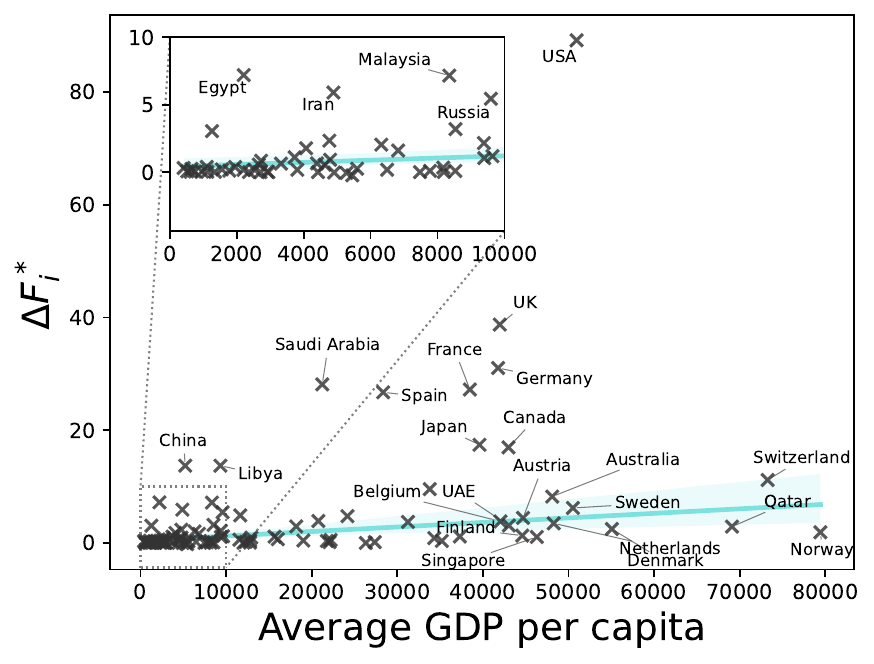}
    \caption{\textbf{Change in funding to study the target Countries as a function of average GDP per Capita per source country}. The moderate correlation (Spearman rank correlation of 0.55) suggests that economic factors may play a significant role in explaining the observed increase in foreign funding across these countries. Notably, high-income countries (HICs) such as the USA, UK, France, Germany, Spain, and Saudi Arabia provided increased funding after the Arab Spring, deviating from broader global patterns.}
    \label{SI-funding_vs_gdp}
\end{figure}

\begin{figure}
    \centering
    \includegraphics[width=0.7\linewidth]{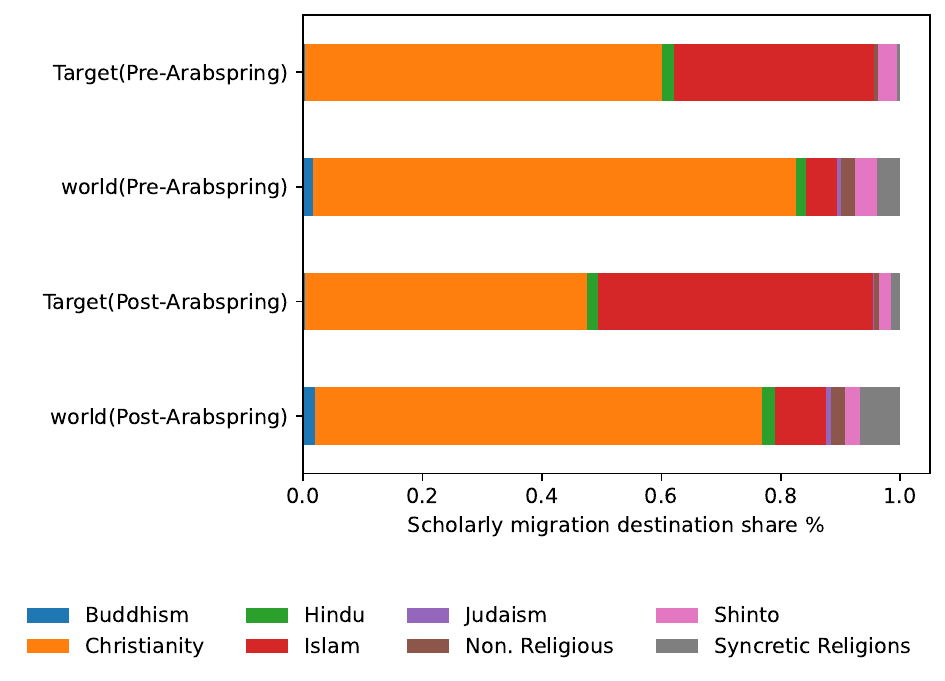}
\caption{\textbf{Scholarly migration patterns by predominant religion of destination.} 
Comparison of migration trends before and after the Arab Spring, focusing on target countries versus global patterns. Researchers from target countries exhibit a higher tendency to migrate to Islamic nations, a trend that has intensified post-Arab Spring.}
\label{SI-Religion_and_migration}

\end{figure}

\begin{figure}
    \centering
    \includegraphics[width=0.7\linewidth]{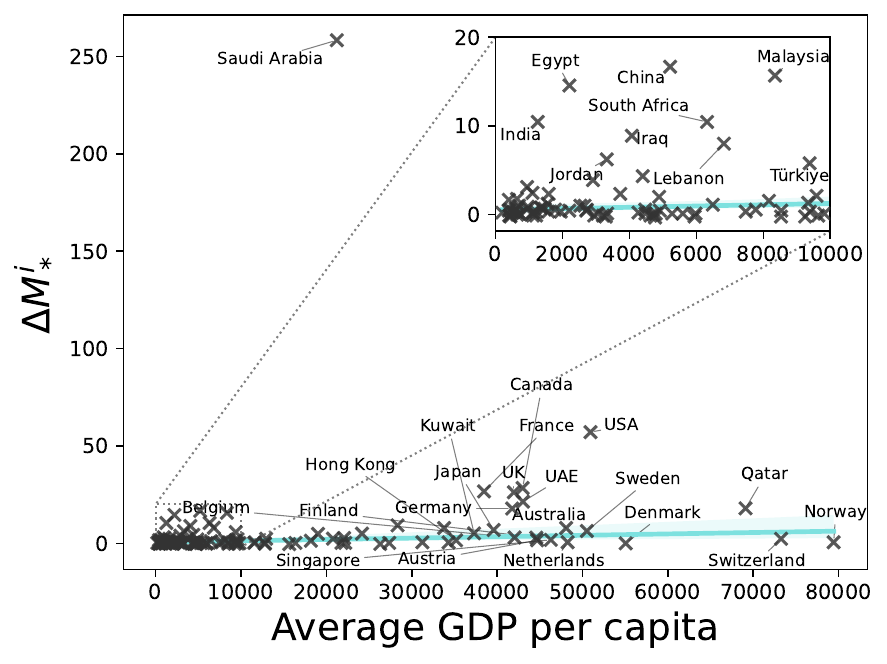}
\caption{\textbf{Scholarly migration from target Countries relative to destination countries' GDP per capita.} 
Despite its strong economy, Saudi Arabia attracted more researchers than the global trend suggests. The correlation between destination countries' average GDP per capita and scholarly migration is moderate (Spearman's \( \rho = 0.33 \), \( p < 0.0001 \)).}
\label{SI-migration_vs_gdp}

\end{figure}

\newpage
\begin{figure}
    \centering
    \includegraphics[width=0.7\linewidth]{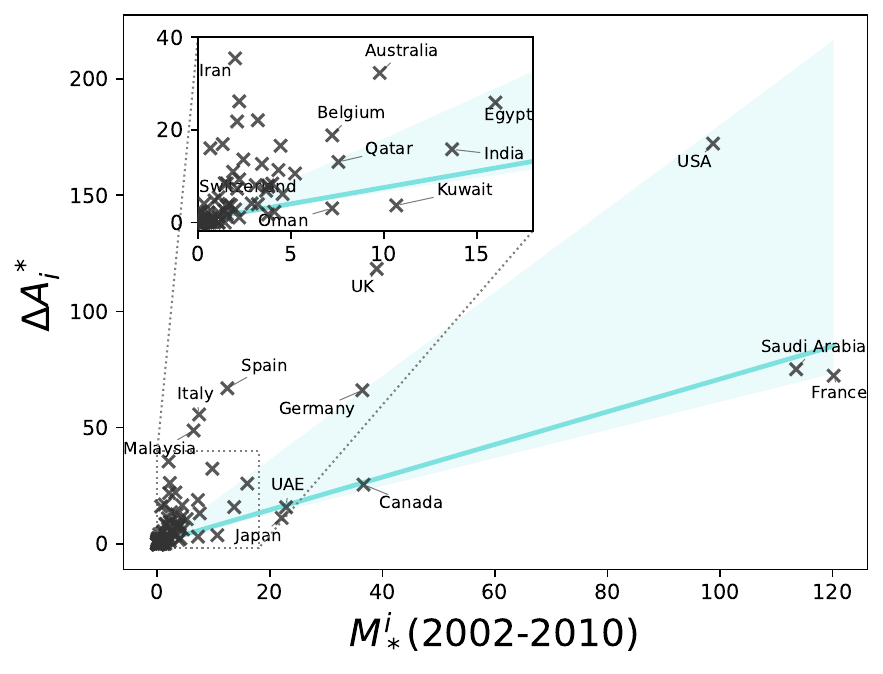}
\caption{\textbf{Change in scholarly attention to target countries as a function of the pre–Arab Spring yearly average of researcher migration from those countries.}  
A strong positive association is observed (Spearman’s \( \rho = 0.745 \), \( p < 0.0001 \); Pearson’s \( \rho = 0.781 \), \( p < 0.0001 \)). Countries such as the USA, Saudi Arabia, and France stand out with both high pre–Arab Spring migration levels and large increases in scholarly attention, while others (e.g., Australia, Iran, Egypt, and India) show disproportionately high attention increases relative to their smaller migration baselines. Most other countries cluster near the origin, reflecting limited migration flows and modest shifts in attention.}
\label{SI-pre_SA_migration_vs_attention}

\end{figure}

\begin{figure}
    \centering
    \includegraphics[width=\linewidth]{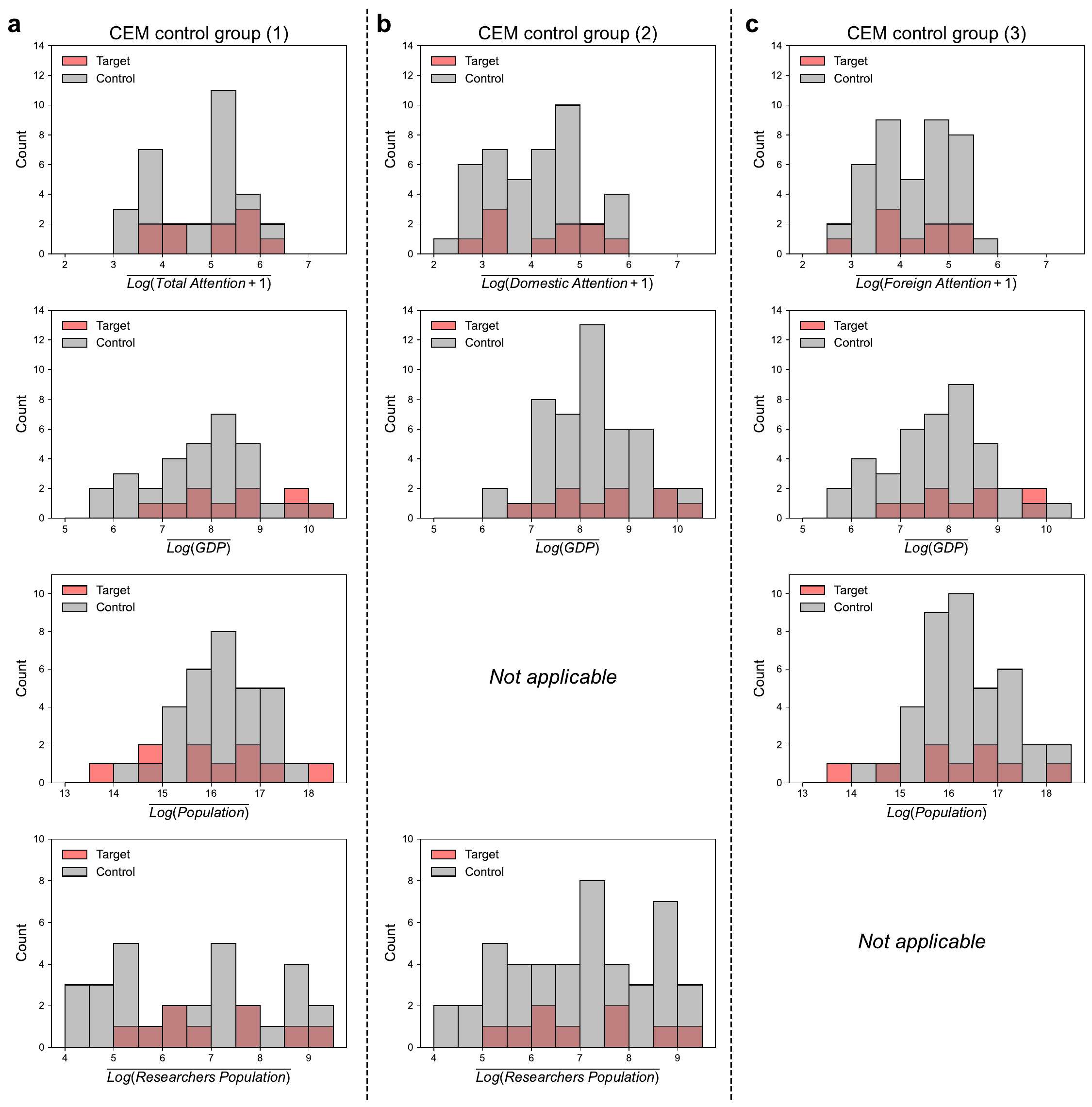}
\caption{\textbf{Distribution of covariates across treated and control units under coarsened exact matching (CEM).}  
Each column corresponds to one CEM control group specification: (a) total attention, (b) domestic attention, and (c) foreign attention. Rows display the distributions of pre-Arab Spring average log-transformed covariates: attention, GDP, population, and researcher population. Gray bars indicate the matched control units, while red bars show the treated (target) units. The overlapping shapes suggest that the CEM procedure achieves a reasonable balance across key covariates, though some discrepancies remain. Empty panels are marked as “Not applicable,” reflecting variables not included for CEM matching.}

\label{SI-covariate_check}

\end{figure}

\clearpage

\begin{figure}
    \centering
\begin{tcolorbox}[colback=gray!20, colframe=black, boxrule=1pt, sharp corners]
\begin{adjustbox}{minipage=\linewidth,scale=0.9, center}
TITLE-ABS-KEY("*arab spring*" OR "*arab-spring*" OR "*arab uprising*" OR "*arab-uprising*" OR "2011 revolution*" OR "2011 uprising*" OR "middle east uprising*" OR (("uprising*" OR "civil unrest*" OR "protests" OR "revolution*") AND ("arab" OR "middle east" OR "north africa"))) AND PUBYEAR\(\geq\)2002 AND PUBYEAR\(\leq\)2019 AND ( LIMIT-TO ( DOCTYPE , "ar" ) ) AND ( LIMIT-TO ( LANGUAGE , "English" ) )
\end{adjustbox}
\end{tcolorbox}
    \caption{\textbf{Scopus query for Arab Spring-related research} This query retrieves academic literature on the Arab Spring from Scopus-indexed sources, focusing on articles in English between 2002 and 2019. In addition to keywords directly related to the Arab Spring, we included broader terms such as "civil unrest" and "protest," combined with regional identifiers like "North Africa" and "Middle East" to ensure comprehensive coverage.}
    \label{SI-arabspring_query}
\end{figure}

\clearpage
\bibliographystyle{naturemag}
\bibliography{refs}

\end{document}